\documentclass[11pt]{article}
\usepackage{latexsym}
\usepackage{chapterbib}
\usepackage{graphics,graphicx}
\usepackage{amssymb,amsmath,amsfonts,amstext}
\usepackage{hyperref}
\usepackage[bbgreekl]{mathbbol}
\usepackage{color,xcolor}
\usepackage{hyperref}
\usepackage{geometry}  
\usepackage{braket}
\usepackage{cancel}
\usepackage[utf8]{inputenc}

\def\){\right)}
\def\({\left( }
\def\]{\right] }
\def\[{\left[ }

\def\NO{\nonumber}

\newcommand{\be}{\begin{equation}}
\newcommand{\ee}{\end{equation}}

\def\bea{\begin{eqnarray}}
\def\eea{\end{eqnarray}}

\def\bal#1\eal{\begin{align}#1\end{align}}

\def\bald{\begin{aligned}}
\def\eald{\end{aligned}}

\def\bsub{\begin{subequations}}
\def\esub{\end{subequations}}

\def\beqx{\begin{displaymath}}
\def\eeqx{\end{displaymath}}

\newcommand{\bmat}{\left(\begin{array}}
\newcommand{\emat}{\end{array}\right)}

\def\half{\frac{1}{2}}

\def\a{\alpha}
\def\b{\beta}
\def\c{\chi}
\def\d{\delta}
\def\e{\epsilon}
\def\f{\phi}
\def\g{\gamma}
\def\h{\eta}

\def\k{\kappa}
\def\l{\lambda}
\def\m{\mu}
\def\n{\nu}
\def\o{\omega}
    
\def\p{\pi}

    \def\th{\theta}
\def\r{\rho}
\def\s{\sigma}

\def\x{\xi}
\def\z{\zeta}
\def\D{\Delta}
\def\F{\Phi}

\def\L{\Lambda}
\def\O{\Omega}
    
\def\P{\Pi}

\def\S{\Sigma}

\def\ve{\varepsilon}
\def\vr{\varrho}

\def\vf{\varphi}

\def\ca{{\cal A}}

\def\cc{{\cal C}}

\def\cf{{\cal F}}
\def\cg{{\cal G}}
\def\ch{{\cal H}}

\def\ck{{\cal K}}
\def\cl{{\cal L}}
\def\cm{{\cal M}}
\def\cn{{\cal N}}
\def\co{{\cal O}}

\def\cq{{\cal Q}}
\def\car{{\cal R}}
\def\cs{{\cal S}}

\def\cz{{\cal Z}}

\def\bb#1{\ensuremath{\mathbb{#1}}} 

\def\bo{{\raise-.3ex\hbox{\large$\Box$}}}               
\def\pa{\partial}                                       
\def\face{{\raise.2ex\hbox{$\displaystyle \bigodot$}\mskip-2.2mu \llap {$\ddot
        \smile$}}}                                   
\def\>{\rangle}                                      
\def\<{\langle}                                      

\def\wt#1{\widetilde{#1}}                            
\def\Hat#1{\widehat{#1}}                             
\def\leftrightarrowfill{$\mathsurround=0pt \mathord\leftarrow \mkern-6mu
        \cleaders\hbox{$\mkern-2mu \mathord- \mkern-2mu$}\hfill
        \mkern-6mu \mathord\rightarrow$}        
\def\dvec#1{\vbox{\ialign{##\crcr
        \leftrightarrowfill\crcr\noalign{\kern-1pt\nointerlineskip}
        $\hfil\displaystyle{#1}\hfil$\crcr}}}           

\def\-{\hphantom{-}}

\begin{document}

\begin{titlepage}

\pagestyle{empty}

\begin{flushright}
 {\small SISSA 04/2016/FISI\\ UPR-1277-T}
\end{flushright}
\vskip1.5in

\begin{center}
\textbf{\Large Black hole thermodynamics from a variational principle:\\
Asymptotically conical backgrounds}
\end{center}
\vskip0.2in

\begin{center}
{\large Ok Song An$^{a,b,c,}$\footnote{\href{mailto: oan@sissa.it}{\tt oan@sissa.it}}\thinspace , 
Mirjam Cveti\v c$^{d,e,}$\footnote{\href{mailto: cvetic@physics.upenn.edu}{\tt cvetic@physics.upenn.edu}}\thinspace ,   
Ioannis Papadimitriou$^{a,}$\footnote{\href{mailto: ipapadim@sissa.it}{\tt ioannis.papadimitriou@sissa.it}}}
\end{center}
\vskip0.2in

\begin{center}
{\small {$^{a}$}SISSA and INFN - Sezione di Trieste, Via Bonomea 265,\\
34136 Trieste, Italy}\\ \vskip0.1in
{\small {$^{b}$}Department of Physics, Kim Il Sung University,\\ Ryongnam Dong, TaeSong District, Pyongyang, DPR Korea}\\ \vskip0.1in
{\small {$^{c}$}ICTP, Strada Costiera 11, 34014 Trieste, Italy}\\ \vskip0.1in
{\small {$^{d}$}Department of Physics and Astronomy, University of Pennsylvania,\\ Philadelphia, PA 19104-6396, USA}\\ \vskip0.1in
{\small {$^{e}$}Center for Applied Mathematics and Theoretical Physics,\\
University of Maribor, SI2000 Maribor, Slovenia}
\end{center}
\vskip0.2in

\begin{abstract}

The variational problem of gravity theories is directly related to black hole thermodynamics. For asymptotically locally AdS backgrounds it is known that holographic renormalization results in a variational principle in terms of equivalence classes of boundary data under the local asymptotic symmetries of the theory, which automatically leads to finite conserved charges satisfying the first law of thermodynamics. We show that this connection holds well beyond asymptotically AdS black holes. In particular, we formulate the variational problem for $\mathcal{N}=2$ STU supergravity in four dimensions with boundary conditions corresponding to those obeyed by the so called `subtracted geometries'. We show that such boundary conditions can be imposed covariantly in terms of a set of asymptotic second class constraints, and we derive the appropriate boundary terms that render the variational problem well posed in two different duality frames of the STU model. This allows us to define finite conserved charges associated with any asymptotic Killing vector and to demonstrate that these charges satisfy the Smarr formula and the first law of thermodynamics. Moreover, by uplifting the theory to five dimensions and then reducing on a 2-sphere, we provide a precise map between the thermodynamic observables of the subtracted geometries and those of the BTZ black hole. Surface terms play a crucial role in this identification.

\end{abstract}

\end{titlepage}

\tableofcontents
\addtocontents{toc}{\protect\setcounter{tocdepth}{3}}
\renewcommand{\theequation}{\arabic{section}.\arabic{equation}}

\section{Introduction}
\setcounter{equation}{0}

Although asymptotically flat or (anti) de Sitter (A)dS backgrounds  have been studied extensively in (super)gravity and string theory, solutions that are asymptotically supported by matter fields have attracted attention relatively recently. Such backgrounds range from flux vacua in string theory to holographic backgrounds dual to supersymmetric quantum field theories (QFTs) \cite{Klebanov:2000hb,Maldacena:2000yy} and non-relativistic systems \cite{Son:2008ye,Balasubramanian:2008dm,Kachru:2008yh,Taylor:2008tg}, to name a few. Understanding the macroscopic properties of black holes with such exotic asymptotics is not only essential in order to address questions of stability and uniqueness, but also a first step towards their microscopic description.     

Thermodynamic quantities such as the black hole entropy or temperature are not sensitive to the asymptotic structure of spacetime, since they are intrinsically connected with the horizon, but observables like conserved charges and the free energy depend heavily on the spacetime asymptotics. This is particularly important 
for backgrounds that are asymptotically supported by matter fields because the conserved pre-symplectic current that gives rise to conserved charges receives contributions from the matter fields \cite{Lee1990,Wald:1999wa}. As a result, the usual methods for computing the conserved charges, such as Komar integrals, often do not work. Moreover, the large distance divergences that plague the free energy and the conserved charges cannot be remedied by techniques such as background subtraction, since it is not always easy, or even possible, to find a suitable background with the same asymptotics. The main motivation behind this paper is  
addressing these difficulties using a general and systematic approach that does not rely on the specific details of the theory or its asymptotic solutions, even though we will demonstrate the general methodology using a concrete example.  

The backgrounds we are going to consider were originally obtained from generic multi-charge asymptotically flat black holes in four \cite{Cvetic:1995kv,Cvetic:1996kv,Chong:2004na,Chow:2013tia,Chow:2014cca} and five dimensions \cite{Cvetic:1996xz} through a procedure dubbed `subtraction' \cite{Cvetic:2011hp,Cvetic:2011dn}. The subtraction procedure consists in excising the asymptotic flat region away from the black hole by modifying the warp factor of the solution, in such a way that the scalar wave equation acquires a manifest $SL(2,\bb R)\times SL(2,\bb R)$ conformal symmetry. 
This leaves the near-horizon region intact, but the resulting background is asymptotically conical \cite{Cvetic:2012tr}. Moreover, it is not necessarily a solution of the original equations of motion.  

It was later realized that the subtracted geometries are solutions \cite{Cvetic:2011dn,Cvetic:2012tr} of the STU model in four dimensions, an $\cn=2$ supergravity theory coupled to three vector multiplets \cite{Duff:1995sm}. The STU model can be obtained from a $T^2$ reduction of minimal supergravity coupled to a tensor multiplet in six dimensions. In particular, the bosonic action is obtained from the reduction of 6-dimensional bosonic string theory 
\be\label{6D-action}
2\k_6 ^2\cl_6=R\star 1-\frac12\star d\f\wedge d\f-\frac12 e^{-\sqrt{2}\f}\star F_{(3)}\wedge F_{(3)},
\ee
where $F_{(3)}=d B_{(2)}$, and then dualizing the 4-dimensional 2-form to an axion. The resulting 4-dimensional theory has an $O(2,2)\simeq SL(2,\bb R)\times SL(2,\bb R)$ global symmetry, which is enhanced to $SL(2,\bb R)^3$ on-shell, when electric-magnetic S-duality transformations are included \cite{Chong:2004na}.

In \cite{Cvetic:2012tr} it was shown that subtracted geometries correspond to a scaling limit of the general non-extremal 4-charge rotating asymptotically flat black hole solutions of the STU model \cite{Cvetic:1996kv,Chong:2004na}, with all four $U(1)$ gauge fields electrically sourced. In \cite{Virmani:2012kw}, starting with the same non-extremal asymptotically flat black holes, but in a frame where only one gauge field is electrically sourced while the remaining three are magnetically sourced, it was shown that the subtracted geometries can also be obtained by Harrison transformations, a solution generating technique exploiting the hidden $SO(4,4)$ symmetry of the STU model upon reduction on a Killing vector \cite{Chong:2004na}. General interpolating solutions between asymptotically flat black holes in four and five dimensions and their subtracted geometry counterparts were subsequently constructed in \cite{Cvetic:2013vqi} by extending these techniques. 

When uplifted to five dimensions the subtracted geometries become a BTZ$\times S^2$ background, with the 2-sphere fibered over the BTZ black hole  \cite{Cvetic:2011dn,Cvetic:2014ina}, which makes manifest the origin of the $SL(2,\bb R)\times SL(2,\bb R)$ symmetry of the wave equation. Using this connection with the BTZ black hole, \cite{Baggio:2012db} showed that the parameters that need to be tuned in order to interpolate between the asymptotically flat black holes and the subtracted geometries correspond to the couplings of irrelevant scalar operators in the two-dimensional conformal field theory (CFT) at the boundary of the asymptotically AdS$_3$ factor of the five-dimensional geometry.   

The thermodynamics of asymptotically conical black holes were first studied in  \cite{Cvetic:2014nta}. In the present work we emphasize the importance of the variational problem in black hole thermodynamics. Using lessons from asymptotically AdS backgrounds \cite{Papadimitriou:2005ii}, we show that a well posed variational problem automatically ensures that all thermodynamic observables are finite and satisfy the first law of thermodynamics. This relegates the problem of seeking the correct definition of conserved charges in backgrounds with new exotic asymptotics to that of properly formulating the variational principle, which in non-compact spaces can be achieved through the following algorithmic procedure: 
\begin{enumerate}

\item[i)] Firstly, the integration constants parameterizing solutions of the equations of motion must be separated into `normalizable' and `non-normalizable' modes. A complete set of modes parameterizes the symplectic space of asymptotic solutions.  
Normalizable modes are free to vary in the variational problem, while non-normalizable modes should be kept fixed.

\item[ii)] Secondly, the non-normalizable modes  are not determined uniquely, but only up to transformations induced by the local symmetries of the bulk theory, such as bulk diffeomorphisms and gauge transformations. Hence, what should be kept fixed in the variational problem is in fact the equivalence class of non-normalizable modes under such transformations \cite{Papadimitriou:2005ii}.

\item[iii)] Formulating the variational problem in terms of equivalence classes of non-normalizable modes requires the addition of a covariant boundary term, $S_{\rm ct}$, to the bulk action, which can be determined by solving asymptotically the radial Hamilton-Jacobi equation \cite{Papadimitriou:2010as}. Since radial translations are part of the local bulk symmetries, formulating the variational problem in terms of equivalence classes ensures that the total action is independent of the radial coordinate, and hence free of long-distance divergences.    

\item[iv)] Finally, besides determining the boundary term $S_{\rm ct}$, the first class constraints of the radial Hamiltonian formulation of the bulk dynamics also lead to conserved charges associated with Killing vectors. The canonical transformation generated by the boundary term $S_{\rm ct}$ `renormalizes' the phase space variables such that these charges are independent of the radial cutoff, and hence finite. These charges automatically satisfy the first law of thermodynamics, with all normalizable modes treated as free parameters and the non-normalizable modes allowed to vary only within the equivalence class under local bulk symmetries.

\end{enumerate}
  
Although this algorithm originates in the AdS/CFT correspondence and holographic renormalization \cite{Henningson:1998gx,Balasubramanian:1999re,deBoer:1999tgo,Kraus:1999di,deHaro:2000vlm,Bianchi:2001de,Bianchi:2001kw,Martelli:2002sp,Skenderis:2002wp,Papadimitriou:2004ap}, it is in principle applicable to any gravity theory, including the subtracted geometries we consider here. However, in this case we find that there are two additional complications, both of which have been encountered before in a holographic context. The first complication arises from the fact that subtracted geometries are obtained as solutions of the STU model provided certain conditions are imposed on the non-normalizable modes. For example, it was shown in \cite{Baggio:2012db} that certain modes (interpreted as couplings of irrelevant scalar operators in the dual CFT$_2$) need to be turned off in the asymptotically flat solutions in order to obtain  the subtracted geometries. We show that all conditions among the non-normalizable modes required to obtain the subtracted geometries can be expressed as covariant second class constraints on the phase space of the theory. This is directly analogous to the way Lifshitz asymptotics were imposed in \cite{Chemissany:2014xsa}. The presence of asymptotic second class constraints in these backgrounds is crucial for being able to solve the radial Hamilton-Jacobi equation and to obtain the necessary boundary term $S_{\rm ct}$.     

The second complication concerns specifically the duality frame in which the STU model was presented in e.g. \cite{Cvetic:2012tr,Cvetic:2014nta}. In this particular  frame, one of the $U(1)$ gauge fields supporting the subtracted geometries  asymptotically dominates the stress tensor, which is reminiscent of fields in asymptotically AdS space that are holographically dual to an irrelevant operator. The variational problem for such fields is known to involve additional subtleties \cite{vanRees:2011fr}, which we also encounter in this specific duality frame of the STU model. We address these subtleties by first formulating the variational problem in a different duality frame and then dualizing to the frame where these complications arise. Remarkably, the form of the boundary term that we obtain through this procedure is exactly of the same form as the boundary term for fields dual to irrelevant operators in asymptotically AdS backgrounds.         

It should be emphasized that our analysis of the variational problem and the derivation of the necessary boundary terms does not assume or imply any holographic duality for asymptotically conical black holes in four dimensions. Nevertheless, subtracted geometries possess a hidden (spontaneously broken) $SL(2,\bb R)\times SL(2,\bb R)\times SO(3)$ symmetry which can be traced to the fact that they uplift to an $S^2$ fibered over a three-dimensional BTZ black hole in five dimensions \cite{Cvetic:2011dn,Cvetic:2012tr}.
The most obvious candidate for a holographic dual, therefore, would be a two-dimensional CFT at the boundary of the asymptotically AdS$_3$ factor of the 5D uplift \cite{Baggio:2012db}. However, if a holographic dual to asymptotically conical backgrounds in four dimensions exists, it is likely that its Hilbert space overlaps with that of the two-dimensional CFT only partially. In particular, we show that the variational problems in four and five dimensions are not fully compatible in the sense that not all asymptotically conical backgrounds uplift to asymptotically AdS$_3\times S ^2$ solutions in five dimensions, and conversely, not all asymptotically AdS$_3\times S ^2$ backgrounds reduce to solutions of the STU model. This is because turning on generic sources on the boundary of AdS$_3$ leads to Kaluza-Klein modes in four dimensions that are not captured by the STU model, while certain modes that are free in the four-dimensional variational problem must be frozen or quantized in order for the solutions to be uplifted to 5D. Although we do not pursue a holographic understanding of the subtracted geometries in the present work, elucidating the relation between the four and five-dimensional variational problems allows us to find a precise map between the thermodynamics of asymptotically conical black holes in four dimensions and that of the BTZ black hole.         

The rest of the paper is organized as follows. We begin with a review of the STU model and the relevant truncations in two distinct duality frames in section \ref{model}, paying particular attention to the surface terms that arise from the dualization procedure. In section \ref{conical} we reparameterize the subtracted geometries in a way that simplifies the separation of the parameters into boundary conditions and dynamical modes that are allowed to vary independently in the variational problem. Moreover, by analyzing the asymptotic symmetries we identify the equivalence classes of boundary conditions in terms of which the variational problem must be formulated. Section \ref{hr} contains the main technical results of the paper. After arguing that the subtraction procedure, i.e. excising the asymptotically flat region in order to zoom into the conical asymptotics of the subtracted geometries, can be implemented in terms of covariant second class constraints on the phase space of the STU model, we derive the covariant boundary terms required in order to formulate the variational problem in terms of equivalence classes of boundary conditions under the asymptotic symmetries. The same boundary terms ensure that the on-shell action is free of long-distance divergences and allows us to construct finite conserved charges associated with any asymptotic Killing vector. In section \ref{thermo} we evaluate explicitly these conserved charges for the subtracted geometries and demonstrate that they satisfy the Smarr formula and the first law of black hole thermodynamics. Section \ref{5&3D} discusses the uplift of the STU model to five dimensions and the Kaluza-Klein reduction  of the resulting theory on the internal $S^2$ to three dimensions, which relates the subtracted geometries to the BTZ black hole. By keeping track of all surface terms arising in this sequence of uplifts and reductions, we provide a precise map between the thermodynamics of the subtracted geometries and that of the BTZ black hole. We end with some concluding remarks in section \ref{conclusion}. Some technical details are presented in two appendices.

\section{The STU model and duality frames}
\label{model}
\setcounter{equation}{0}

In this section we review the bosonic sector of the 2-charge truncation of the STU model that is relevant for describing the subtracted geometries. We will do so in the duality frame discussed in \cite{Cvetic:2012tr}, where both charges are electric, as well as in the one used in \cite{Virmani:2012kw}, where there is one electric and one magnetic charge. We will refer to these frames as `electric' and `magnetic' respectively. As it will become clear from the subsequent analysis, in order to compare the thermodynamics in the two frames, it is necessary to keep track of surface terms introduced by the duality transformations. 

\subsection{Magnetic frame}

The bosonic Lagrangian of the STU model in the duality frame used in \cite{Virmani:2012kw} is given by
\bal\label{STU}
2\k_4^2\cl_{4}=\, & R\star1-\frac{1}{2}\star d\h_a\wedge d\h_a-\frac{1}{2}e^{2\h_{a}}\star d\c^{a}\wedge d\c^{a}\NO\\
&-\frac{1}{2}e^{-\h_{0}}\star F^{0}\wedge F^{0}-\frac{1}{2}e^{2\h_{a}-\h_{0}}\star(F^{a}+\c^{a}F^{0})\wedge(F^{a}+\c^{a}F^{0})\NO\\
& +\frac{1}{2}C_{abc}\c^{a}F^{b}\wedge F^{c}+\frac{1}{2}C_{abc}\c^{a}\c^{b}F^{0}\wedge F^{c}+\frac{1}{6}C_{abc}\c^{a}\c^{b}\c^{c}F^{0}\wedge F^{0},
\eal
where $\h_{a}$ ($a=1,2,3$) are dilaton fields and $\h_{0}=\sum_{a=1}^{3}\h_{a}$. The symbol $C_{abc}$ is pairwise symmetric with $C_{123}=1$ and zero otherwise. The Kaluza-Klein ansatz for obtaining this action from the 6-dimensional action \eqref{6D-action} is given explicitly in \cite{Virmani:2012kw}. This frame possesses an explicit triality symmetry, exchanging the three gauge fields $A^a$, the three dilatons $\h^a$ and the three axions $\c^a$. In this frame, the subtracted geometries source all three gauge fields $A^a$ magnetically, while $A^0$ is electrically sourced. Moreover, holographic renormalization turns out to be much more straightforward in this frame compared with the electric frame. 

In order to describe the subtracted geometries it suffices to consider a truncation of \eqref{STU}, corresponding to setting $\h_1=\h_2=\h_3\equiv\h$, $\c_1=\c_2=\c_3\equiv\c$, and $A^1=A^2=A^3\equiv A$. The resulting action can be written in the $\s$-model form
\be\label{STU-reduced}
S_4=\frac{1}{2\k^2_4}\int_\cm\text{d}^4 \mathbf{x} \sqrt{-g}\left(R[g]-\frac12\cg_{IJ}\pa_{\mu}\vf^I\pa^{\mu}\vf^J-\cz_{\L\S}F^\L_{\m\n}F^{\S\m\n}-\car_{\L\S}\e^{\mu\nu\rho\s}F^\L_{\mu\nu}F^\S_{\rho\s}\right)+S_{\rm GH},
\ee
where
\be\label{GH}
S_{\rm GH}=\frac{1}{2\k^2_4}\int_{\pa\cm} \text{d}^3 \mathbf{x} \sqrt{-\g}\;2K,
\ee
is the standard Gibbons-Hawking \cite{Gibbons:1976ue} term and we have defined the doublets  
\be
\vf^I=\(\begin{matrix}
\h \\
\c
\end{matrix}\),\qquad 
A^\L=\(\begin{matrix}
A^0 \\
A
\end{matrix}\),\qquad I=1,2,\quad \L=1,2,
\ee
as well as the $2\times 2$ matrices
\be
\cg_{IJ}=\(\begin{matrix}
3 & 0 \\
0 & 3e^{2\h}
\end{matrix}\),\qquad 
\cz_{\L\S}=\frac14\(\begin{matrix}
e^{-3\h}+3e^{-\h}\c^2 & 3e^{-\h}\c\\
3e^{-\h}\c  & 3e^{-\h}
\end{matrix}\),\qquad 
\car_{\L\S}=\frac14\(\begin{matrix}
\c^3 & \frac32\c^2\\
 \frac32\c^2& 3\c
\end{matrix}\).
\ee
As usual, $\e_{\m\n\r\s}=\sqrt{-g}\;\ve_{\m\n\r\s}$ denotes the totally antisymmetric Levi-Civita tensor, where $\ve_{\m\n\r\s}=\pm1$ is the Levi-Civita symbol. Throughout this paper we choose the orientation in $\cm$ so that $\ve_{rt\th\f}=1$.  We note in passing that the Lagrangian \eqref{STU-reduced} is invariant under the global symmetry transformation
\be\label{global-symm-m}
e^\h\to \m^2 e^\h, \quad\c\to\m^{-2}\c,\quad  A^0\to\m^3 A^0,\quad A\to \m A,\quad ds^2\to ds^2,
\ee 
where $\m$ is an arbitrary non-zero constant parameter.

\subsection{Electric frame}

The STU model in the duality frame in which the subtracted geometries are presented in \cite{Cvetic:2012tr} can be obtained from \eqref{STU-reduced} by dualizing the gauge field $A$.\footnote{\label{comparison} Notice that the duality frame in eq.~(1) of \cite{Cvetic:2012tr} is not the one in which the solutions are given in that paper. As mentioned above eq.~(3), two of the gauge fields in (1) are dualized in the solutions discussed. The corresponding action, which was not given explicitly in \cite{Cvetic:2012tr}, can be obtained from the magnetic frame action \eqref{STU-reduced} here by first implementing the field redefinitions $A\to -A$ and $\c\to-\c$ and then dualizing $A$ as we describe here. The resulting action differs by a few signs from our electric frame action \eqref{STU-reduced-e}. } Following \cite{Chow:2014cca}, we dualize $A$ by introducing a Lagrange multiplier, $\wt A$, imposing the Bianchi identity $dF=0$, and consider the action 
\be\label{STU-dualization}
\wt S_4=S_4+\frac{1}{2\k_4^2}\int_\cm 3\wt A \wedge dF=S_4+\frac{1}{2\k_4^2}\int_\cm 3\wt F \wedge F-\frac{3}{2\k_4^2}\int_{\pa\cm}\wt A\wedge F+\frac{3}{2\k_4^2}\int_{\ch_+}\wt A\wedge F.
\ee
The factor of 3 is a convention, corresponding to a choice of normalization for $\wt A$, chosen such that the resulting electric frame model agrees with the one in \cite{Cvetic:2012tr}. The term added to $S_4$ vanishes on-shell and so the on-shell values of $\wt S_4$ and $S_4$ coincide. The total derivative term that leads to surface contributions from the boundary, $\pa\cm$, and the outer horizon, $\ch_+$, is crucial for comparing the physics in the electric and magnetic frames. As we will discuss later on, this surface term is also the reason behind the subtleties of holographic renormalization in the electric frame.  
 
Integrating out $F$ in \eqref{STU-dualization} we obtain
\be\label{F}
F_{\m\n}=-(4\c ^2+e^{-2\h })^{-1}\(\frac12\e_{\m\n\r\s}\;e^{-\h }(\wt F-\c^2F^0)^{\r\s}+2\c \wt F_{\m\n}+\c (2\c ^2+e^{-2\h })F^0_{\m\n}\).
\ee 
Inserting this expression for $F$ in \eqref{STU-dualization} leads to the electric frame action 
\bal
\wt S_4 
&=\frac{1}{2\k^2_4}\int_\cm \(R\star \mathbf 1-\frac 32 \star d\h \wedge d\h -\frac 32 e^{2\h }\star d\c \wedge d\c -\half e^{-3\h } \star F^0 \wedge F^0\right.\NO\\
&\left.\hskip0.5in-\frac32\frac{e^{-\h}}{(4\c ^2+e^{-2\h })}\star(\wt F-\c^2F^0)\wedge (\wt F-\c ^2 F^0)\right.\NO\\
&\left.\hskip0.5in-\frac{\c}{(4\c ^2+e^{-2\h })}\[3\wt F\wedge\wt F+3(2\c^2+e^{-2\h})\wt F\wedge F^0-\c^2(\c^2+e^{-2\h})F^0\wedge F^0\]\)\NO\\
&\hskip0.5in -\frac{3}{2\k_4^2}\int_{\pa\cm}\wt A\wedge F+\frac{3}{2\k_4^2}\int_{\ch_+}\wt A\wedge F+S_{\rm GH}.
\eal
As in the magnetic frame, it is convenient to write the bulk part of the action in $\s$-model form as 
\bal\label{STU-reduced-e}
\wt S_4&=\frac{1}{2\k^2_4}\int_\cm\text{d}^4 \mathbf{x} \sqrt{-g}\left(R[g]-\frac12\cg_{IJ}\pa_{\mu}\vf^I\pa^{\mu}\vf^J-\wt \cz_{\L\S}\wt F^\L_{\m\n}\wt F^{\S\m\n}-\wt\car_{\L\S}\e^{\mu\nu\rho\s}\wt F^\L_{\mu\nu}\wt F^\S_{\rho\s}\right)\NO\\
&\hskip1.5in-\frac{3}{2\k_4^2}\int_{\pa\cm}\wt A\wedge F+\frac{3}{2\k_4^2}\int_{\ch_+}\wt A\wedge F+S_{\rm GH},
\eal
where we have defined 
\bal\label{e-frame}
& \wt A^\L=\(\begin{matrix}
A^0 \\
\wt A
\end{matrix}\),\qquad \wt \cz_{\L\S}=\frac14\(\begin{matrix}
e^{-3\h}+\frac{3e^{-\h}\c^4}{4\c^2+e^{-2\h}} & -\frac{3 e^{-\h}\c^2}{4\c^2+e ^{-2\h}}\\
-\frac{3 e^{-\h}\c^2}{4\c^2+e ^{-2\h}}  & \frac{3e^{-\h}}{4\c^2+e^{-2\h}}
\end{matrix}\), \NO\\ 
&\wt\car_{\L\S}=\frac{\c}{4(4\c ^2+e^{-2\h })}\(\begin{matrix}
\c^2(\c^2+e^{-2\h}) & -\frac{3}{2}(2\c^2+e^{-2\h})\\
 -\frac{3}{2}(2\c^2+e^{-2\h})& -3
\end{matrix}\).
\eal
As in the magnetic frame, the action \eqref{STU-reduced-e} is invariant under the global symmetry transformation
\be\label{global-symm-e}
e^\h\to \m^2 e^\h, \quad\c\to\m^{-2}\c,\quad  A^0\to\m^3 A^0,\quad \wt A\to \m^{-1}\wt A,\quad ds^2\to ds^2.
\ee

\section{Asymptotically conical backgrounds}
\label{conical}
\setcounter{equation}{0}

The general rotating subtracted geometry backgrounds are solutions of the equations of motion following from the action \eqref{STU-reduced} or \eqref{STU-reduced-e} and take the form \cite{Cvetic:2012tr,Virmani:2012kw}\footnote{In order to compare this background with the expressions given in eqs. (24) and (25) of \cite{Cvetic:2012tr}, one should take into account the field redefinition $A\to -A$, $\c\to -\c$, before the dualization of $A$, as mentioned in footnote \ref{comparison}, and add a constant pure gauge term. Moreover, there is a typo in eq. (25) of \cite{Cvetic:2012tr}: the term $2m\P_s^2\cos^2\th d\bar t$ should be replaced by $2m\P_s(\P_s-\P_c)\cos^2\th d\bar t$.}
\bal\label{subtracted}
ds^2&=\frac{\sqrt{\D}}{X}d\bar r^2-\frac{G}{\sqrt{\D}}(d\bar t+\ca)^2+\sqrt{\D}\(d\th^2+\frac{X}{G}\sin^2\th d\bar\phi^2\),\NO\\
e^\h &=\frac{(2m)^2}{\sqrt{\D}},\qquad \c=\frac{a\(\P_c-\P_s\)}{2m}\cos\th,\NO\\
A^0 &=\frac{(2m)^4a\(\P_c-\P_s\)}{\D}\sin^2\th d\bar \f+\frac{(2ma)^2\cos^2\th\(\P_c-\P_s\)^2+(2m)^4\P_c\P_s}{\(\P_c^2-\P_s^2\)\D}d\bar t,\NO\\
A & =\frac{2m\cos\th}{\D}\(\[\D-(2ma)^2(\Pi_c-\Pi_s)^2\sin^2\th\]d\bar\f-2ma\(2m\P_s+\bar r(\P_c-\P_s)\)d\bar t\,\rule{0.0cm}{0.35cm}\),\NO\\
\wt A & =-\frac{1}{2m}\(\bar r-m-\frac{(2ma)^2(\P_c-\P_s)}{(2m)^3(\P_c+\P_s)}\)d\bar t
+\frac{(2ma)^2(\P_c-\P_s)[2m\P_s+\bar r(\P_c-\P_s)]\cos^2\th}{2m\D}d\bar t\NO\\
&\hskip1.4in +a(\P_c-\P_s)\sin^2\th\(1+\frac{(2ma)^2(\P_c-\P_s)^2\cos^2\th}{\D}\)d\bar\f,
\eal
where
\bal
& X=\bar r^2-2m\bar r+a^2, \qquad G= X-a^2 \sin^2\th,\qquad \ca=\frac{2ma}{G} \((\Pi_c-\Pi_s)\bar r+2m\Pi_s\)\sin^2\th d\bar \phi,\NO\\
& \D=(2m)^3(\Pi_c^2-\Pi_s^2)\bar r+(2m)^4\Pi_s^2-(2ma)^2(\Pi_c-\Pi_s)^2\cos^2\th,
\eal
and $\P_c$, $\P_s$, $a$ and $m$ are parameters of the solution. 

In order to study the thermodynamics of these backgrounds it is necessary to identify which parameters are fixed by the boundary conditions in the variational problem. A full analysis of the variational problem for the actions \eqref{STU-reduced} or \eqref{STU-reduced-e} requires knowledge of the general asymptotic solutions and is beyond the scope of the present paper. However, we can consider the variational problem within the class of stationary solutions \eqref{subtracted}. To this end, it is convenient to reparameterize these backgrounds by means of a suitable coordinate transformation, accompanied by a relabeling of the free parameters. In particular, we introduce the new coordinates
\bal
\label{coordinate-trans}
\ell^4r &=(2m)^3(\Pi_c^2-\Pi_s^2)\bar r+(2m)^4\Pi_s^2-(2ma)^2(\Pi_c-\Pi_s)^2,\NO\\
\frac{k}{\ell^3}t &= \frac{1}{(2m)^3(\Pi_c^2-\Pi_s^2)}\bar t,\qquad \f = \bar\phi-\frac{2ma(\Pi_c-\Pi_s)}{(2m)^3(\Pi_c^2-\Pi_s^2)}\bar t,
\eal
where $\ell$ and $k$ are additional non-zero parameters, whose role will become clear shortly. Moreover, we define the new parameters
\bal
\ell^4r_\pm & =(2m)^3m(\Pi_c^2+\Pi_s^2)-(2ma)^2(\Pi_c-\Pi_s)^2\pm\sqrt{m^2-a^2}(2m)^3(\Pi_c^2-\Pi_s^2),\NO\\
\ell^3\o &=2ma(\Pi_c-\Pi_s),\qquad B=2m,
\eal
which can be inverted in order to express the old parameters in terms of the new ones, namely
\bal
\P_{c,s}&=\frac{\ell^2}{B^2}\(\frac12\(\sqrt{r_+}+\sqrt{r_-}\)\pm\sqrt{\ell^2\o^2+\frac14\(\sqrt{r_+}-\sqrt{r_-}\)^2}\),\NO\\
a&=\frac{B\ell\o}{2\sqrt{\ell^2\o^2+\frac14\(\sqrt{r_+}-\sqrt{r_-}\)^2}},\qquad
m=B/2.
\eal
Rewriting the background \eqref{subtracted} in terms of the new coordinates and parameters we obtain\footnote{\label{continuity} Since these solutions carry non-zero magnetic charge, the gauge potential $A$ must be defined in the north ($\th<\p/2$) and south hemispheres respectively as \cite{Copsey:2005se,Chow:2013gba}, $A_{\rm north}=A-Bd\f$ and $A_{\rm south}=A+Bd\f$, where $A$ is the expression given in \eqref{subtracted-simple}. }
\bal
\label{subtracted-simple}
e^\h&=\frac{B^2/\ell^2}{\sqrt{r+\ell^2\o^2\sin^2\th}},\qquad \c=\frac{\ell^3\o}{B^2}\cos\th,\NO\\
A^0&=\frac{B^3/\ell^3}{r+\ell^2\o^2\sin^2\th}\(\sqrt{r_{+}r_{-}}\;kdt+\ell^2\o\sin^2\th d\phi\),\NO\\
A&=\frac{B\cos\th}{r+\ell^2\o^2\sin^2\th}\(-\o\sqrt{r_{+}r_{-}}\;kdt+rd\phi\),\NO\\
\wt A & = -\frac{\ell}{B}\(r-\frac12(r_++r_-)\)kdt+\frac{\o\ell^3}{B}\cos^2\th\(\frac{\o\sqrt{r_+r_-}\; kdt-rd\f}{r+\o^2\ell^2\sin^2\th}\)+\frac{\o\ell^3}{B} d\f,\NO\\ 
ds^2&=\sqrt{r+\ell^2\o^2\sin^2\th}\(\frac{\ell^2dr^2}{(r-r_{-})(r-r_{+})}-\frac{(r-r_{-})(r-r_{+})}{r}k^2dt^2+\ell^2 d\th^2\)\NO\\
&\hskip2.6in+\frac{\ell^2 r\sin^2\th}{\sqrt{r+\ell^2\o^2\sin^2\th}}\(d\f-\frac{\o\sqrt{r_+ r_-}}{r}k dt\)^2.
\eal

Several comments are in order here. Firstly, the two parameters $r_\pm$ are the locations of the outer and inner horizons respectively, and clearly correspond to normalizable perturbations. A straightforward calculation shows that $\o$ is also a normalizable mode. We will explicitly confirm this later on by showing that the long-distance divergences of the on-shell action are independent of $\o$. Setting the normalizable parameters to zero we arrive at the background
\bal
\label{subtracted-vacuum}
e^\h&=\frac{B^2/\ell^2}{\sqrt{r}},\qquad \c= 0,\qquad A^0= 0,\qquad
A= B\cos\th d\phi,\qquad
\wt A = -\frac{\ell}{B}rkdt,\NO\\
ds^2&= \sqrt{r}\(\ell^2\frac{dr^2}{r^2}-r k^2dt^2+\ell^2d\th^2+\ell^2\sin^2\th d\f^2\),
\eal
which we shall consider as the vacuum solution. The fact that the background \eqref{subtracted-vacuum} is singular does not pose any difficulty since it should only be viewed as an asymptotic solution that helps us to properly formulate the variational problem. Changing the radial coordinate to $\vr=\ell r^{1/4}$, the vacuum metric becomes  
\be
ds^2= 4^2d\vr^2-\(\frac{\vr}{\ell}\)^6 k^2dt^2+\vr^2\(d\th^2+\sin^2\th d\f^2\),
\ee
which is a special case of the conical metrics discussed in \cite{Cvetic:2012tr}. Different conical geometries are supported by different matter fields. Although we focus on the specific conical backgrounds obtained as solutions of the STU model here, we expect that our analysis, modified accordingly for the different matter sectors, applies to general asymptotically conical backgrounds.

The asymptotic structure of (stationary) conical backgrounds is parameterized by the three non-zero constants $B$, $\ell$ and $k$. In the most restricted version of the variational problem, these three parameters should be kept fixed. However, there is a 2-parameter family of deformations of these boundary data still leading to a well posed variational problem, as we now explain. The first deformation corresponds to the transformation of the boundary data induced by reparameterizations of the radial coordinate. Namely, under the bulk diffeomorphism  
\be\label{PBH}
r\to \l^{-4} r,\quad \l> 0,
\ee
the boundary parameters transform as 
\be\label{source-scaling}
k\to \l^{3}k,\quad \ell\to \l\ell,\quad B\to B.
\ee
This transformation is a direct analogue of the so called Penrose-Brown-Henneaux (PBH) diffeomorphisms in asymptotically AdS backgrounds \cite{Imbimbo:1999bj}, which induce a Weyl transformation on the boundary sources. The PBH  diffeomorphisms imply that the bulk fields do not induce boundary fields, but only a {\em conformal structure}, that is boundary fields up to Weyl transformations \cite{Skenderis:2002wp}. This dictates that the variational problem must be formulated in terms of conformal classes rather than representatives of the conformal class \cite{Papadimitriou:2005ii}. In the case of subtracted geometries, variations of the boundary parameters of the form   
\be\label{var1}
\d_1 k=3\e_1 k,\quad \d_1\ell =\e_1\ell, \quad \d_1 B=0,
\ee
correspond to motion within the equivalence class (anisotropic conformal class) defined by the transformation \eqref{source-scaling}, and therefore lead to a well posed variational problem. 

A second deformation of the boundary data that leads to a well posed variational problem is 
\be\label{var2}
\d_2 k=0,\quad \d_2\ell =\e_2\ell, \quad \d_2 B=\e_2 B.
\ee
To understand this transformation, one must realize that the parameters $B$ and $\ell$ do not correspond to independent modes, but rather only the ratio $B/\ell$, which can be identified with the source of the dilaton $\h$. In particular, keeping $B/\ell$ fixed ensures that the variational problem is the same in all frames of the form
\be
ds_\a^2=e^{\a\h}ds^2,
\ee
for some $\a$, which will be important for the uplift of the conical backgrounds to five dimensions. The significance of the parameter $B$ is twofold. It corresponds to the background magnetic field in the magnetic frame and variations of $B$ are equivalent to the global symmetry transformation \eqref{global-symm-m} or \eqref{global-symm-e} of the bulk Lagrangian. Moreover, as we will discuss in the next section, it enters in the covariant asymptotic second class constraints imposing conical boundary conditions. The transformation \eqref{var2} is a variation of $B$ combined with a bulk diffeomorphism in order to keep the modes $k$ and $B/\ell$ fixed. The relevant bulk diffeomorphism is a rescaling of the radial coordinate of the form \eqref{PBH}, accompanied by a rescaling $t\to \l^{3}t$ of the time coordinate.

\section{Boundary counterterms and renormalized conserved charges}
\label{hr}
\setcounter{equation}{0}

The first law of black hole thermodynamics is directly related to the variational problem and the boundary conditions imposed on the solutions of the equations of motion. As we briefly reviewed in the previous section, in non-compact spaces, where the geodesic distance to the boundary is infinite, the bulk fields induce only an equivalence class of boundary fields, which implies that the variational problem must be formulated in terms of equivalence classes of boundary data, with different elements of the equivalence class related by radial reparameterizations. In order to formulate the variational problem in terms of equivalence classes of boundary data one must add a specific boundary term, $S_{\rm ct}=-\cs_o$, to the bulk action, where $\cs_o$ is a certain asymptotic solution of the radial Hamilton-Jacobi equation, which we discuss in appendix \ref{ham}. For asymptotically locally AdS spacetimes, this boundary term is identical to the boundary counterterms derived by the method of holographic renormalization \cite{Henningson:1998gx,Balasubramanian:1999re,deBoer:1999tgo,Kraus:1999di,deHaro:2000vlm,Bianchi:2001de,Bianchi:2001kw,Martelli:2002sp,Skenderis:2002wp,Papadimitriou:2004ap}, which are designed to render the on-shell action free of large-distance divergences. In particular, demanding that the variational problem be formulated in terms of equivalence classes (conformal classes in the case of AdS) of boundary data cures all pathologies related to the long-distance divergences of asymptotically AdS spacetimes, leading to a finite on-shell action and conserved charges that obey the first law and the Smarr formula of black hole thermodynamics \cite{Papadimitriou:2005ii}. This observation, however, goes beyond asymptotically AdS backgrounds. Provided a suitable asymptotic solution $\cs_o$ of the radial Hamilton-Jacobi equation can be found, one can perform a canonical transformation of the form
\be
\(\f^\a,\;\p_\b\)\to \(\f^\a,\;\P_\b=\p_\b-\frac{\d\cs_o}{\d\f^\b}\),
\ee      
such that the product $\f^\a\P_\a$ depends only on the equivalence class of boundary data. This in turn implies that formulating the variational problem in terms of the symplectic variables $(\f^\a,\; \P_\b)$ ensures that it be well posed \cite{Papadimitriou:2010as}.        

This analysis of the variational problem presumes that the induced fields $\f^\a$ on a slice of constant radial coordinate are independent variables, or equivalently, the boundary data induced from the bulk fields are unconstrained. However, this may not be the case. Imposing conditions on the boundary data leads to different asymptotic structures and accordingly different boundary conditions. A typical example is the case of asymptotically Lifshitz backgrounds \cite{Kachru:2008yh,Taylor:2008tg} (see \cite{Taylor:2015glc} for a recent review), where non-relativistic boundary conditions are imposed on a fully diffeomorphic bulk theory \cite{Chemissany:2014xsa}. The conditions imposed on the boundary data correspond to asymptotic second class constraints of the form 
\be\label{constraint}
\cc\(\f^\a\)\approx 0,
\ee
in the radial Hamiltonian formulation of the bulk dynamics. As a result, the asymptotic solution $\cs_o$ of the Hamilton-Jacobi equation that should be added as a boundary term may not be unique anymore, since it can be written in different ways, all related to each other by means of the constraints \eqref{constraint}. It should be emphasized that the potential ambiguity in the boundary counterterms arising due to the presence of asymptotic second class constraints is not related to the ambiguity that is commonly referred to as `scheme dependence' in the context of the AdS/CFT correspondence \cite{Bianchi:2001kw}. The latter is an ambiguity in the {\em finite} part of the solution $\cs_o$, and it exists independently of the presence of second class constraints. On the contrary, the potential ambiguity resulting from the presence of second class constraints may affect both the divergent and finite parts of $\cs_o$. As we will see below, in order to obtain asymptotically conical backgrounds from the STU model one must impose certain asymptotic second class constraints, which play a crucial role in the understanding of the variational problem. A subset of these second class constraints corresponds to turning off the modes that, if non-zero, would lead to an asymptotically Minkowski background. As such, the asymptotic second class constraints constitute a covariant way of turning off the couplings of the irrelevant scalar operators identified in \cite{Baggio:2012db}, or implementing the original subtraction procedure.       

After covariantizing the definition of asymptotically conical backgrounds in the STU model by introducing a set of covariant second class constraints, we will determine the boundary terms required in order to render the variational problem well posed, both in the magnetic and electric frames. This will allow us to define finite conserved charges associated with asymptotic Killing vectors, which will be used in section \ref{thermo} in order to prove the first law of thermodynamics for asymptotically conical black holes. We will first consider the magnetic frame because the electric frame presents additional subtleties, which can be easily addressed once the variational problem in the magnetic frame is understood. Since the boundary term we must determine in order to render the variational problem well posed is a solution of the radial Hamilton-Jacobi equation, the analysis in this section relies heavily on the radial Hamiltonian formulation of the bulk dynamics discussed in detail in appendix \ref{ham}. In particular, we will work in the coordinate system \eqref{ADM} and gauge-fix the Lagrange multipliers as     
\be\label{gf}
N=\(r+\ell^2\o^2\sin^2\th\)^{1/4},\quad N_i=0, \quad a^\L=\wt a^\L=0.
\ee

\subsection{Magnetic frame}

Even though we have not determined the most general asymptotic solutions of the equations of motion compatible with conical boundary conditions in the present work, we do need a covariant definition of asymptotically conical backgrounds in order to determine the appropriate boundary term that renders the variational problem well posed. It turns out that the stationary solutions \eqref{subtracted-simple} are sufficiently general in order to provide a minimal set of covariant second class constraints, which can be deduced from the asymptotic form \eqref{subtracted-vacuum} of conical backgrounds. In the magnetic frame they take the form 
\be
\label{asymptotic-constraints}
F_{ij}F^{ij}\approx \frac{2}{B^2}e^{2\h},\quad
R_{ij}[\g]\approx e^{-\h}F_{ik} F_j{}^{k},\quad
2R_{ij}[\g]R^{ij}[\g]\approx R[\g]^2, 
\ee
where the $\approx$ symbol indicates that these constraints should be imposed only asymptotically, i.e. they should be understood as conditions on non-normalizable modes only. Qualitatively, these covariant and gauge-invariant second class constraints play exactly the same role as the second class constraints imposing Lifshitz asymptotics \cite{Chemissany:2014xsa}. 

The fact that we have been able to determine the constraints \eqref{asymptotic-constraints} in covariant form ensures that the boundary term we will compute below renders the variational problem well posed for general asymptotically conical backgrounds -- not merely the stationary solutions \eqref{subtracted-simple}. Moreover, this boundary term can be used together with the first order equations \eqref{flow} to obtain the general asymptotic form of conical backgrounds, but we leave this analysis for future work.

\subsubsection{Boundary counterterms}

The general procedure for determining the solution $\cs_o$ of the Hamilton-Jacobi equation, and hence the boundary counterterms, is the following. Given the leading asymptotic form of the background, the first order equations \eqref{flow} are integrated asymptotically in order to obtain the leading asymptotic form of $\cs_o$. Inserting this leading solution in the Hamilton-Jacobi equation, one sets up a recursive procedure that systematically determines all subleading corrections that contribute to the long-distance divergences. Luckily, for asymptotically conical backgrounds in four dimensions, integrating the first order equations \eqref{flow} using the leading asymptotic form of the background determines all divergent terms, and so there is no need for solving the Hamilton-Jacobi recursively.      

In order to integrate the first order equations \eqref{flow} we observe that  
the radial coordinate $u$ in \eqref{ADM} is related to the coordinate $r$ in \eqref{subtracted-simple} as 
\be\label{coords}
du=\frac{\ell dr}{\sqrt{(r-r_+)(r-r_-)}}\sim\ell dr/r,\qquad \pa_u=\ell^{-1}\sqrt{(r-r_+)(r-r_-)}\sim\ell^{-1}r\pa_r.
\ee
Using these relations, together with the asymptotic form \eqref{subtracted-vacuum} of the conical backgrounds, we seek to express the radial derivatives of the induced fields as covariant functions of the induced fields. In particular, focusing on the three first order equations that are relevant for our computation, it is not difficult to see that to leading order asymptotically one can write 
\be\label{velocities}
\frac1N\dot{\g}_{ij}\sim \frac{e^{\h/2}}{B} \(\frac32\g_{ij}- B^2e^{-2\h} F_{ik} F_j{}^k\),\quad
\frac1N\dot\h \sim -\frac{e^{\h/2}}{2B},\quad
\frac1N\dot A^0_i \sim \frac B2 e^{3\h} D_j\(e^{-7\h/2} F^{0j}{}_i\). 
\ee
Notice that the first two expressions are not unique since they can be written in alternative ways using the constraints \eqref{asymptotic-constraints}. Taking into account these expressions, as well as the freedom resulting from the constraints, we conclude that the leading asymptotic form of the solution of the Hamilton-Jacobi equation takes the form
\be\label{HJ-sol} 
\cs=\frac{1}{\k_4^2}\int\text{d}^3 \mathbf{x}\sqrt{-\g}\;\frac{1}{B}e^{\h/2}\(a_1+a_2 B^2 e^{-\h}R[\g]+a_3 B^2 e^{-2\h} F_{ij}F^{ij}+a_4B^2 e^{-4\h}F^0_{ij}F^{0ij}+\cdots\),
\ee
where $a_1$, $a_2$, $a_3$ and $a_4$ are unspecified constants and the ellipses stand for subleading terms.  The functional derivatives of this asymptotic  solution take the form
\bsub
\label{S-derivatives}
\bal
\frac{\d \cs}{\d\g_{ij}}&=\frac{\sqrt{-\g}}{\k_4^2}\frac{1}{B}e^{\h/2}\(\frac12\g^{ij}\(a_1+a_2 B^2 e^{-\h}R[\g]+a_3 B^2 e^{-2\h} F_{kl}F^{kl}+a_4B^2 e^{-4\h}F^0_{kl}F^{0kl}+\cdots\)\right.\NO\\
&\left.\hskip1.2in+a_2 B^2e^{-\h}\(-R^{ij}+\frac14\pa^i\h\pa^j\h-\frac12 D^iD^j\h-\frac14\g^{ij}\pa_k\h\pa^k\h+\frac12\g^{ij}\square_\g\h\)\right.\NO\\
&\left.\hskip2.5in-2a_3 B^2e^{-2\h} F^{ik} F^j{}_k-2a_4B^2e^{-4\h}F^{0ik}F^{0j}{}_k+\cdots\rule{0cm}{0.5cm}\)\NO\\
&\sim \frac{\sqrt{-\g}}{\k_4^2}\frac{1}{B}e^{\h/2}\(\frac12(a_1+2a_2+2a_3)\g^{ij}-(a_2+2a_3)B^2e^{-2\h} F^{ik}F^j{}_k+\cdots\),\\
\frac{\d \cs}{\d\h}&=\frac{\sqrt{-\g}}{\k_4^2}\frac{1}{B}e^{\h/2}\frac12\(a_1-a_2B^2e^{-\h}R[\g]-3a_3B^2 e^{-2\h} F_{ij} F^{ij}-7a_4B^2 e^{-4\h}F^0_{ij}F^{0ij}+\cdots\),\NO\\
&\sim \frac{\sqrt{-\g}}{\k_4^2}\frac{1}{B}e^{\h/2}\frac12\(a_1-2a_2-6a_3\)+\cdots,\\
\frac{\d \cs}{\d A^0_{i}}&=-\frac{\sqrt{-\g}}{\k_4^2}4B a_4 D_j\(e^{-7\h/2}F^{0ji}\)+\cdots,
\eal
\esub
where the symbol $\sim$ indicates that we have used the constraints \eqref{asymptotic-constraints} and only kept the leading terms. Inserting these in the first order equations \eqref{flow} and comparing with \eqref{velocities} leads to the set of algebraic equations
\be
a_1-2a_2-6a_3=\frac32,\quad a_2+2a_3=-\frac14,\quad a_4=\frac{1}{16},
\ee
which admit the one-parameter family of solutions 
\be
a_1=1-\a/4,\quad a_2=(\a-1)/4,\quad a_3=-\a/8,\quad a_4=1/16,
\ee 
where $\a$ is unconstrained. One can readily check that \eqref{HJ-sol}, with these values for $a_1$, $a_2$, $a_3$ and $a_4$, satisfies the Hamilton-Jacobi equations asymptotically for any value of the parameter $\a$. 

As we shall see momentarily, for any $\a$, this asymptotic solution suffices to remove all long-distance divergences of the on-shell action and renders the variational problem well posed on the space of equivalence classes of boundary data.
We have therefore determined that a complete set of boundary counterterms for the variational problem in the magnetic frame is 
\be\label{counterterms} 
S_{\rm ct}=-\frac{1}{\k_4^2}\int\text{d}^3 \mathbf{x}\sqrt{-\g}\;\frac{B}{4}e^{\h/2}\(\frac{4-\a}{B^2}+(\a-1)e^{-\h}R[\g]-\frac{\a}{2} e^{-2\h}F_{ij} F^{ij}+\frac{1}{4}e^{-4\h}F^0_{ij}F^{0ij}\). 
\ee
As we mentioned at the beginning of this section, the freedom to choose the value of the parameter $\a$ does not correspond to a choice of scheme. Instead, it is a direct consequence of the presence of the second class constraints \eqref{asymptotic-constraints}. The scheme dependence corresponds to the freedom to include additional finite local terms, which do not affect the divergent part of the solution. Later on we will consider situations where additional conditions on the variational problem require a specific value for $\a$, or particular finite counterterms.

\subsubsection{The variational problem}

Given the counterterms $S_{\rm ct}$ and following standard terminology in the context of the AdS/CFT duality, we define the `renormalized' on-shell action in the magnetic frame as the sum of the on-shell action \eqref{STU-reduced} and the counterterms \eqref{counterterms}, with the regulating surface $\S_u$ removed. Namely,
\be\label{ren-action-m}
S_{\rm ren}=\lim_{r\to\infty}\(S_4+S_{\rm ct}\).
\ee
The boundary counterterms ensure that this limit exists and its value is computed in appendix \ref{4Daction-evaluation}. 

A generic variation of the renormalized on-shell action takes the form
\be\label{variation-m}
\d S_{\rm ren}=\lim_{r\to\infty}\int\text{d}^3 \mathbf{x}\(\P^{ij}\d\g_{ij}+\P^i_\L\d A^\L_i+\P_I\d\vf^I\),
\ee
where the renormalized canonical momenta are given by
\be\label{ren-momenta-def}
\P^{ij}=\p^{ij}+\frac{\d S_{\rm ct}}{\d\g_{ij}},\quad 
\P^{i}_\L=\p^{i}_\L+\frac{\d S_{\rm ct}}{\d A^\L_{i}},\quad
\P_I=\p_I+\frac{\d S_{\rm ct}}{\d\vf^I}.
\ee
Inserting the asymptotic form of the backgrounds \eqref{subtracted-simple} into the definitions \eqref{momenta} of the canonical momenta and in the functional derivatives \eqref{S-derivatives} we obtain 
\bsub
\label{ren-momenta}
\bal 
&\P ^t_{\phantom{t}t} \sim -\frac{k\ell}{2\k_4^2}\(\frac14(r_+ + r_-)+\frac{\a-2}{8}\ell^2\o^2(1+3\cos 2\th) \)\sin\th,\quad \P ^\f_{\phantom{t}t}  \sim-\frac{k^2\ell\o}{2\k_4^2} \sqrt{r_+ r_-} \sin\th ,\\
&\P ^\th_{\phantom{t}\th} \sim \frac{k\ell^3\o ^2}{16\k_4^2}((2-5\a )\cos 2\th +2-3\a)\sin\th ,\quad
\P ^\f_{\phantom{t}\f} \sim-\frac{k\ell^3\o ^2}{16\k_4^2}((5\a -4)\cos 2\th+3\a)\sin\th ,\\
&\P^{0t}  \sim-\frac{1}{2\k_4^2}\frac{\ell^4}{B^3}\sin\th\(\sqrt{r_+ r_-}+3\o ^2\ell^2\cos^2\th  \), \quad
\P^{0\f}  \sim-\frac{1}{2\k_4^2}\frac{k\o\ell^4}{2B^3}(r_+ + r_-)\sin\th,\\
&\P^{t}  \sim -\frac{1}{2\k_4^2}\frac{3\o\ell^3}{B}\sin 2\th, \quad
\P^{\f}  \sim -\frac{1}{2\k_4^2}\frac{2\a k\o ^2\ell^3}{B}\sin 2\th, \\
&\P_\h  \sim-\frac{1}{2\k_4^2}\frac{k\ell}{8}\sin\th \(6(r_+ + r_-)+\ell^2\o ^2 ((13\a -18)\cos 2\th+7\a -6) \),
\eal
\esub
with all other components vanishing identically.

Finally, we can use these expressions to evaluate the variation \eqref{variation-m} of the renormalized action in terms of boundary data. To this end we need to perform the integration over $\th$ and remember that the magnetic potential $A$ is not globally defined, as we pointed out in footnote \ref{continuity}. In particular, taking
$A_{\rm north}\sim B(\cos\th -1) d\f$ and $A_{\rm south}\sim B(\cos\th +1) d\f$ we get 
\be\label{mode-var}
\d S_{\rm ren}=-\frac{1}{2\k_4^2}\int \text{d}t \text{d}\f\;(r_++r_-)k\ell\d\log\(kB^3/\ell^3\),
\ee
independently of the value of the parameter $\a$. Note that the combination $kB^3/\ell^3$ of boundary data is the unique invariant under both the equivalence class transformation \eqref{var1} and the transformation \eqref{var2}. We have therefore demonstrated that by adding the counterterms \eqref{counterterms} to the bulk action, the variational problem is formulated in terms of equivalence classes of boundary data under the transformations \eqref{var1} and \eqref{var2}. This is an explicit demonstration of the general result that formulating the variational problem in terms of equivalence classes of boundary data under radial reparameterizations is achieved via the same canonical transformation that renders the on-shell action finite. As we will now demonstrate, the same boundary terms ensure the finiteness of the conserved charges, as well as the validity of the first law of thermodynamics.

\subsubsection{Conserved charges}

Let us now consider conserved charges associated with local conserved currents. This includes electric charges, as well as conserved quantities related to  asymptotic Killing vectors. Magnetic charges do not fall in this category, but they can be described in this language in the electric frame, as we shall see later on. 

In the radial Hamiltonian formulation of the bulk dynamics, the presence of local conserved currents is a direct consequence of the first class constraints $\cf^\L=0$ and $\ch^i=0$ in \eqref{constraints}.\footnote{These constraints can be derived alternatively by applying the general variation \eqref{variation-m} of the renormalized action to $U(1)$ gauge transformations and transverse diffeomorphisms, assuming the invariance of the renormalized action under such transformations. This method will be used in order to derive the conserved charges in the electric frame.} As in the case of asymptotically AdS backgrounds, these constraints lead respectively to conserved electric charges and charges associated with asymptotic Killing vectors.\footnote{In asymptotically locally AdS spaces, the Hamiltonian constraint $\ch=0$ can be used in order to construct conserved charges associated with {\em conformal} Killing vectors of the boundary data \cite{Papadimitriou:2005ii}. For asymptotically conical backgrounds, the Hamiltonian constraint leads to conserved charges associated with asymptotic transverse diffeomorphisms, $\x^i$, that preserve the boundary data up to the equivalence class transformations \eqref{var1}.} In particular, the gauge constraints $\cf^\L=0$ in \eqref{constraints} take the form
\be
D_i\p^i=0, \quad D_i\p^{0i}=0, 
\ee
where $\p^i$ and $\p^{0i}$ are respectively the canonical momenta conjugate to the gauge fields $A_i$ and $A^{0}_i$. Since the boundary counterterms \eqref{counterterms} are gauge invariant, it follows from \eqref{ren-momenta-def} that these conservation laws hold for the corresponding renormalized momenta as well, namely 
\be
D_i\P^i=0, \quad D_i\P^{0i}=0. 
\ee
This implies that the quantities 
\be\label{e-charges-m}
Q_4^{(e)} =-\int_{\pa\cm\cap C}\text{d}^2 \mathbf{x} \, \P^{t},\qquad Q_4^{0(e)} =-\int_{\pa\cm\cap C}\text{d}^2 \mathbf{x} \, \P^{0t},
\ee
where $C$ denotes a Cauchy surface that extends to the boundary $\pa\cm$, are both conserved and finite and correspond to the electric charges associated with these gauge fields.   
 
Similarly, the momentum constraint $\ch^i=0$ in \eqref{constraints}, which   
can be written in explicit form as 
\bal
&-2D_j\pi^{j}_i+\pi_\h\pa_i\h+\pi_\c\pa_i\c+F^0_{ij}\pi^{0j}+ F_{ij}\pi^{j}\NO\\
&+\frac{1}{2\k^2_4}\sqrt{-\g}\;\e^{jkl}\(\c^3F^0_{ ij}F^0_{kl}+\frac32\c^2F^0_{ij} F_{kl}+3\c  F_{ij} F_{kl}+\frac32\c^2 F_{ij}F^0_{kl}\)=0, 
\eal 
leads to finite conserved charges associated with asymptotic Killing vectors. Note that the terms in the second line are independent of the canonical momenta and originate in the parity odd terms in the STU model Lagrangian.\footnote{In the AdS/CFT context these terms are interpreted as a gravitational anomaly in the dual QFT.} However, for asymptotically conical backgrounds of the form \eqref{subtracted-simple} these terms are asymptotically subleading, the most dominant term being
\be
\sqrt{-\g}\e^{jkl}\c F_{ij}F_{kl}=\co(r^{-1}),
\ee 
and so the momentum constraint asymptotically reduces to 
\be
-2D_j\pi^{j}_i+\pi_\h\pa_i\h+\pi_\c\pa_i\c+F^0_{ij}\pi^{0j}+ F_{ij}\pi^{j}\approx 0.
\ee
Since the counterterms \eqref{counterterms} are invariant with respect to diffeomorphisms along the surfaces of constant radial coordinate, it follows from \eqref{ren-momenta-def} that this constraint holds for the renormalized momenta as well, 
\be\label{WI-m}
-2D_j\P^{j}_i+\P_\h\pa_i\h+\P_\c\pa_i\c+F^0_{ij}\P^{0j}+F_{ij}\P^{j}\approx 0.
\ee

Given an asymptotic Killing vector $\z^i$ satisfying the asymptotic conditions 
\be\label{Killing}
\cl_\z\g_{ij}=D_i\z_j+D_j\z_i\approx 0,\quad \cl_\z A^\L_i=\z^j\pa_j A^\L_i+A^\L_j\pa_i\z^j\approx 0,\quad \cl_\z\vf^I=\z^i\pa_i\vf^I\approx 0,
\ee 
the conservation identity \eqref{WI-m} implies that the quantity 
\be\label{conserved-Q} 
\cq[\z]= \int_{\partial\cm\cap C} \text{d}^2 \mathbf{x}\,\(2\P^t_j+\P^{0t}A^0_j+\P^tA_j\)\z^j, 
\ee
is both finite and conserved, i.e. it is independent of the choice of Cauchy surface $C$. However, there are a few subtleties in evaluating these charges. Firstly, Gauss' theorem used to prove conservation for the charges \eqref{conserved-Q} assumes differentiability of the integrand across the equator at the boundary. If the gauge potentials are magnetically sourced, as is the case for $A_i$ in the magnetic frame, then the gauge should be chosen such that $A_i$ is continuous across the equator. In particular, contrary to the variational problem we discussed earlier, the gauge that should be used to evaluate these charges is the one given in \eqref{subtracted-simple}, and not the one discussed in footnote \ref{continuity}. 

Secondly, the charges \eqref{conserved-Q} are not generically invariant under the $U(1)$ gauge transformations $A_i^\L\to A_i^\L+\pa_i\a^\L$. These gauge  transformations though must preserve both the radial gauge \eqref{gf} and the asymptotic Killing conditions \eqref{Killing}. Preserving the radial gauge implies that the gauge parameter must depend only on the transverse coordinates, i.e. $\a^\L(x)$ (see e.g. \cite{Papadimitriou:2005ii}), while respecting the Killing symmetry leads to the condition    
\be\label{gauge-parameter}
\z^i\pa_i\a^\L=\text{constant}. 
\ee
Under such gauge transformations the charges \eqref{conserved-Q} are shifted by the corresponding electric charges \eqref{e-charges-m}. As will become clear in section \ref{thermo}, this compensates a related shift in the electric potential such that the Smarr formula and the first law are gauge invariant. Nevertheless, gauge invariant charges, as well as electric potentials, can be defined if and only if $\left.A^\L_j\z^j\right|_{\pa\cm}=\text{constant}$. However, this is not true in general. 

Finally, another potential ambiguity in the value of the charges \eqref{conserved-Q} arises from the ambiguity in the choice of boundary counterterms used to define the renormalized momenta. In the case of asymptotically conical backgrounds in the magnetic frame, this ambiguity consists in both the value of the parameter $\a$ in \eqref{counterterms}, as well as the possibility of adding extra finite and covariant terms. From the explicit expressions \eqref{ren-momenta} we see that $\a$ does lead to an ambiguity in the renormalized momenta. However, as we will see in section \ref{thermo}, it does not affect the value of physical observables.

The fact that the value of the charges \eqref{conserved-Q} is ambiguous in the precise sense we just discussed does not affect the thermodynamic relations among the charges and the first law, which are unambiguous. In fact, the ambiguity in the definition \eqref{conserved-Q} of the conserved charges allows us to match them to alternative definitions \cite{Hollands:2005wt,Papadimitriou:2005ii}.

\subsection{Electric frame}

We will now repeat the above analysis for the variational problem in the electric frame, emphasizing the differences relative to the magnetic frame. Besides the fact that the electric frame is most commonly used in the literature on subtracted geometries, it is also necessary in order to evaluate the magnetic potential, also known as the `magnetization'. Moreover, the variational problem for asymptotically conical backgrounds in the electric frame presents some new subtleties, from which interesting lessons can be drawn. 

\subsubsection{Boundary counterterms}

By construction, the electric frame action $\wt S_4$ given in \eqref{STU-reduced-e} has the same on-shell value as the magnetic frame action $S_4$ in \eqref{STU-reduced}. Therefore, the boundary counterterms \eqref{counterterms} that were derived for $S_4$ must also render the variational problem for $\wt S_4$ well posed and remove its long-distance divergences. Adding the boundary counterterms \eqref{counterterms} to $\wt S_4$ we get
\be\label{e-frame-action}
\wt S_4+S_{\rm ct}=\wt S_4'+S_{\rm ct}-\frac{3}{2\k_4^2}\int_{\pa\cm}\wt A\wedge F+\frac{3}{2\k_4^2}\int_{\ch_+}\wt A\wedge F,
\ee
where $\wt S_4'$ denotes the $\s$-model part of \eqref{STU-reduced-e} (plus the Gibbons-Hawking term), to which the Hamiltonian analysis of appendix \ref{ham} can be applied. 

As for the bulk part of the action in \eqref{STU-reduced-e}, we need to replace $F_{ij}$ in the boundary terms with the electric gauge field $\wt A_i$ using \eqref{F}, which for the transverse components reduces to the canonical momentum for $\wt A_i$ in \eqref{momenta}, namely  
\be\label{momentum-e}
\wt \pi^i_\L=\frac{\d L}{\d\dot {\wt  A^\L }_i}=-\frac{2}{\k_4^2}\sqrt{-\g}\(N^{-1}\wt \cz_{\L\S}\g^{ij}\dot {\wt A^\S}_j+\wt \car_{\L\S}\e^{ijk}\wt F^\S_{jk}\).
\ee 
Evaluating this expression leads to the identity
\be\label{F-momentum}
F_{ij}=\frac{2\k_4^2}{3}\ve_{ijk}\wt\p^k=\frac{2\k_4^2}{3}\e_{ijk}\Hat{\wt\p}^k,
\ee
where we have defined $\Hat{\wt \p}^i=\wt\p^i/\sqrt{-\g}$. Hence,
\be
\wt S_4+S_{\rm ct}=\wt S_4'+S_{\rm ct}-\int_{\pa \cm }  \text{d}^3 \mathbf{x}\; \wt \p ^i \wt A_i+\frac{3}{2\k_4^2}\int_{\ch_+}\wt A\wedge F,
\ee
where the counterterms are now expressed as
\be\label{counterterms-e}
S_{\rm ct}=-\frac{B}{4\k_4^2}\int\text{d}^3 \mathbf{x}\sqrt{-\g}\;e^{\h/2}\(\frac{4-\a}{B^2}+(\a-1)e^{-\h}R[\g]+\frac{1}{4}e^{-4\h}F^0_{ij}F^{0ij}+\frac{4\a \k_4^4}{9}e^{-2\h}\Hat{\wt\p}^i\Hat{\wt\p}_i\). 
\ee
The renormalized action in the electric frame therefore takes the form
\be\label{ren-action-e}
\wt S_{\rm ren}=\lim_{r\to\infty}\(\wt S_4'+S_{\rm ct}-\int_{\pa \cm }  \text{d}^3 \mathbf{x}\; \wt \p ^i \wt A_i\),
\ee
with $S_{\rm ct}$ given by \eqref{counterterms-e}. Moreover, the asymptotic second class constraints \eqref{asymptotic-constraints} become
\be
\label{asymptotic-constraints-e}
\Hat{\wt\p}^k\Hat{\wt\p}_k\approx -\(\frac{3e^{\h}}{2\k_4^2 B}\)^2,\quad
R_{ij}[\g]\approx -\(\frac{2\k_4^2}{3}\)^2e^{-\h}\(\g_{ij}\Hat{\wt\p}^k\Hat{\wt\p}_k-\Hat{\wt\p}_i\Hat{\wt\p}_j\),\quad
2R_{ij}[\g]R^{ij}[\g]\approx R[\g]^2. 
\ee

Note that the surface term on the horizon in \eqref{e-frame-action} is not part of the action defining the theory in the electric frame, which is why we have not included it in the definition of the renormalized action \eqref{ren-action-e}. The theory is specified by the bulk Lagrangian and the boundary terms on $\pa\cm$, which dictate the variational problem and the boundary conditions. The horizon is a dynamical surface -- not a boundary. This surface term, however, will be essential in section \ref{thermo} for comparing the free energies in the electric and magnetic frames.  

Given that the counterterms $S_{\rm ct}$ render the on-shell action in the magnetic frame finite, the limit $\eqref{ren-action-e}$ is guaranteed to exist: its value differs from the on-shell value of the renormalized action \eqref{ren-action-m} by the surface term on the horizon given in \eqref{e-frame-action}. However, as we will show shortly, it turns out that the variational problem for the renormalized action \eqref{ren-action-e} is only well posed provided  $\a$ takes a specific non-zero value. This value is determined by the term implementing the Legendre transformation in \eqref{ren-action-e}, which has a fixed coefficient. Therefore, even though any value of $\a$ leads to a well posed variational problem in the magnetic frame, a specific value of $\a$ is required for the variational problem in the electric frame. 

Another consequence of the Legendre transform in \eqref{ren-action-e} is that it changes the boundary conditions from Dirichlet, where $\wt A_i$ is kept fixed on the boundary (up to equivalence class transformations), to Neumann, where $\wt \p^i$ is kept fixed. This in turn forces the counterterms to be a function of the canonical momentum, i.e. $S_{\rm ct}[\g,A^0,\Hat{\wt\p},\h,\c]$. An analogous situation arises in asymptotically AdS backgrounds with fields that are dual to irrelevant operators \cite{vanRees:2011fr}. An example that shares many qualitative features with the potential $\wt A_i$ here is a gauge field in AdS$_2$, coupled to appropriate matter \cite{O'Bannon:2015gwa}. From the form of the conical backgrounds \eqref{subtracted-simple} we see that $\wt A_i$ asymptotically dominates the stress tensor as $r\to\infty$ since   
\be
T_{tt} \sim e^\h g^{rr}(\wt F_{rt})^2\sim r,
\ee   
and hence, in this sense, the gauge potential $\wt A_i$ is analogous to bulk fields dual to irrelevant operators in asymptotically AdS spaces. This property is what makes the variational problem and the boundary counterterms in the electric frame more subtle, which is why we found it easier to formulate the variational problem in the magnetic frame first and then translate the result to the electric frame.

\subsubsection{The variational problem}

A generic variation of the renormalized action \eqref{ren-action-e} takes the form
\be\label{variation-e}
\d \wt S_{\rm ren}=\lim_{r\to\infty}\int\text{d}^3 \mathbf{x}\(\wt\P^{ij}\d\g_{ij}+\wt\P^{0i}\d A^{0}_i-\sqrt{-\g}\;\wt A^{\rm ren}_i\d\Hat{\wt\p}^i+\wt\P_I\d\vf^I\),
\ee
where
\be 
\wt\P^{ij}=\wt\p^{ij}-\frac12 \g ^{ij}\wt \p ^k \wt A_k+\left.\frac{\d S_{\rm ct}}{\d\g_{ij}}\right|_{\Hat{\wt\p}},\quad \wt\P^{0i}=\wt\p^{0i}+\frac{\d S_{\rm ct}}{\d A^{0}_{i}},\quad
\wt\P_I=\wt\p_I+\frac{\d S_{\rm ct}}{\d\vf^I},
\ee
and
\be\label{ren-A}
\wt A^{\rm ren}_i=\wt A_i-\frac{1}{\sqrt{-\g}}\left.\frac{\d S_{\rm ct}}{\d\Hat{\wt\p}^i}\right|_{\g},
\ee
are the renormalized canonical variables in the electric frame. It should be emphasized that the functional derivative with respect to $\g_{ij}$ in $\wt\P^{ij}$ is computed keeping $\Hat{\wt\p}^i$ fixed instead of $\wt\p^i$. The term implementing the Legendre transform in \eqref{ren-action-e}, therefore, gives $-\frac12 \g ^{ij}\wt \p ^k \wt A_k$, while    
\bal
&\left.\frac{\d S_{\rm ct}}{\d\g_{ij}}\right|_{\Hat{\wt\p}}=-\frac{\sqrt{-\g }}{\k _4^2}\frac{e^{\h /2}}{B}\Bigg (\frac12 \g ^{ij}\(\frac14+\frac12 B^2 e^{-\h }R[\g]+\frac{1}{16}B^2 e^{-4\h}F^0_{ij}F^{0ij} \)-\frac18 B^2e^{-4\h }F^{0ik}F^{0j}_{\phantom {0j}k}\NO\\
& \hskip1.5in +\frac12B^2e^{-\h}\(-R^{ij}+\frac14\pa^i\h\pa^j\h-\frac12 D^iD^j\h-\frac14\g^{ij}\pa_k\h\pa^k\h+\frac12\g^{ij}\square_\g\h\)\Bigg)\NO\\
&\hskip1.5in -\frac{\k _4^2}{\sqrt{-\g }}\frac{B}{3}e^{-3\h /2}\(\frac{\g ^{ij}}{2}\wt \p ^k \wt \p _k+\wt \p ^i \wt \p ^j  \).
\eal 
Moreover, note that \eqref{momenta} implies that the canonical momenta 
$\wt\p^{ij}$, $\wt\p^{0i}$, and $\wt\p_I$, remain the same as their magnetic frame counterparts. 

What is novel in \eqref{variation-e} from the point of view of holographic renormalization is that the variable that gets renormalized is the induced field $\wt A_i$, according to \eqref{ren-A}, instead of its conjugate momentum. However, as we mentioned earlier, only a specific value of the parameter $\a$ correctly renormalizes $\wt A_i$. In particular, from \eqref{counterterms-e} we get
\be 
\wt A^{\rm ren}_i=\wt A_i+\frac{2\a B\k _4^2}{9}\frac{e^{-3\h /2}}{\sqrt{-\g}}\wt \p_i.
\ee 
On the other hand, from \eqref{momentum-e} and the asymptotic form of $\wt A_i$ in \eqref{subtracted-vacuum} we deduce that asymptotically
\be 
\wt \p ^i \sim -\frac{2}{\k _4 ^2}\sqrt{-\g }\cdot \frac{3}{4B}e^{3\h /2}\g ^{it}\wt A_t.
\ee
It follows that $\wt A^{\rm ren}_i$ has a finite limit as $r\to\infty$ provided  
$\a=3$.

Setting $\a=3$ in \eqref{counterterms-e} and evaluating the renormalized variables on the conical backgrounds \eqref{subtracted-simple} we obtain  
\bsub
\label{ren-momenta-e}
\bal
&\wt\P ^t_{\phantom{t}t}  \sim \frac{k\ell}{2\k_4^2}\sin\th\(-\frac14(r_+ + r_-)+\frac18\ell^2\o^2(11+9\cos 2\th) \), \quad \wt\P ^\f_{\phantom{t}t}  \sim -\frac{1}{2\k_4^2}\frac{k\o\ell}{2}\sqrt{r_+ r_-}\sin\th,\\
&\wt\P ^\th_{\phantom{t}\th}  \sim -\frac{1}{2\k_4^2}\frac{k\ell^3\o ^2}{16}(\sin 3\th -11\sin\th),\quad \wt\P ^\f_{\phantom{t}\f}  \sim\frac{1}{2\k_4^2}\frac{k\ell^3\o ^2}{16}(\sin 3\th+5\sin\th), \\
&\wt\P^{0t}  \sim-\frac{1}{2\k_4^2}\frac{\ell^4}{B^3}\sin\th\(\sqrt{r_+ r_-}+3\o ^2\ell^2\cos^2\th  \), \quad
\wt\P^{0\f}  \sim-\frac{1}{2\k_4^2}\frac{k\o\ell^4}{2B^3}(r_+ + r_-)\sin\th,\\
&\wt A_t^{\rm ren}  \sim \frac{2\o ^2\ell ^3 k}{B}\cos^2\th, \quad \wt A_\f^{\rm ren}  \sim \frac{\o\ell^3\sin^2\th}{B},\\
&\wt\P_\h  \sim -\frac{1}{2\k_4^2}\frac{3k\ell}{8}\sin\th \(2(r_+ + r_-)+\ell^2\o ^2 (5+7\cos 2\th  ) \).
\eal
\esub
Finally, inserting these expressions in \eqref{variation-e} gives
\bal 
\d \wt S_{\rm ren} & =-\frac{1}{2\k_4^2}\int \text{d}t \text{d}\f\;(r_++r_-)k\ell\d\log\(kB^3/\ell^3\),
\eal 
in agreement with the magnetic frame result \eqref{mode-var}. In particular, as in the magnetic frame, the variational problem is well posed in terms of equivalence classes of boundary data under the transformation \eqref{var1}.

\subsubsection{Conserved charges}

The last aspect of the electric frame we need to discuss before we can move on to study the thermodynamics of conical backgrounds is how to define the conserved charges. Focusing again on charges obtained from local conserved currents, the electric charges follow from the conservation laws 
\be
D_i\wt \p^i=0,\qquad D_i\wt \P^{0i}=0. 
\ee 
From \eqref{F-momentum} we see that the first of these expressions is simply the Bianchi identity $d F=0$ and so the corresponding charge is the magnetic charge in the magnetic frame, $Q_4^{(m)}$, while $\wt \P^{0i}$ coincides with the renormalized momentum $\P^{0i}$ in the magnetic frame. Hence,   
\be\label{e-charges-e}
\wt Q_4^{(e)} =-\int_{\pa\cm\cap C}\text{d}^2 \mathbf{x}\, \wt\p^{t}=Q_4^{(m)},\qquad \wt Q_4^{0(e)} =-\int_{\pa\cm\cap C}\text{d}^2 \mathbf{x} \, \wt\P^{0t}=Q_4^{0(e)}.
\ee

Slightly more subtle are conserved charges associated with asymptotic Killing vectors. The easiest way to derive the conservation laws in the electric frame is by considering the variation of the renormalized action under an infinitesimal diffeomorphism, $\x^i$, along the surfaces of constant radial coordinate. Inserting the transformations   
\bal
&\d_\x\g_{ij}=\cl_\x\g_{ij}=D_i\x_j+D_j\x_i,\quad \d_\x\vf^I=\cl_\x\vf^I=\x^i\pa_i\vf^I,\NO\\
&\d_\x A^0_i=\cl_\x A^0_i=\x^j\pa_j A^0_i+A^0_j\pa_i\x^j,\quad
\d_\x\Hat{\wt\p}^i=\cl_\x\Hat{\wt\p}^i=\x^jD_j\Hat{\wt\p}^i-\Hat{\wt\p}^jD_j\x^i,
\eal
under such a diffeomorphism in the general variation \eqref{variation-e} of the renormalized action gives
\bal
\d \wt S_{\rm ren}=&\lim_{r\to\infty}\int\text{d}^3 \mathbf{x}\(2\wt\P^{ij}D_i\x_j+\x^iF^0_{ij}\P^{0j}-\sqrt{-\g}\wt A^{\rm ren}_i\d\Hat{\wt\p}^i+\P_I\x^i\pa_i\vf^I\)\NO\\
=&\lim_{r\to\infty}\int\text{d}^3 \mathbf{x}\;\x^i\(-2D_j\wt\P^{j}_i+F^0_{ij}\P^{0j}-D_i\(\wt A^{\rm ren}_j\wt\p^j\)+\wt\cf_{ij}\wt\p^j+\P_I\pa_i\vf^I\),
\eal
from which we arrive at the conservation identity
\be\label{WI-e}
-2D_j\(\wt\P^{j}_i+\frac12\d^j_i\wt\p^k\wt A^{\rm ren}_k\)+F^0_{ij}\P^{0j}+\wt\cf_{ij}\wt\p^j+\P_I\pa_i\vf^I\approx0.
\ee

An asymptotic Killing vector, $\z^i$, in the electric frame satisfies the same conditions \eqref{Killing} as in the magnetic frame, except that 
the asymptotic form of the background is now specified in terms of $\Hat{\wt\p}^i$ instead of $A_i$ and so the condition $\cl_\z A_i=\z^j\pa_j A_i+A_j\pa_i\z^j\approx 0$ in the magnetic frame should be replaced with 
\be
\cl_\z\Hat{\wt\p^i}=\z^jD_j\Hat{\wt\p}^i-\Hat{\wt\p}^jD_j\z^i\approx 0.
\ee
With this crucial modification in the definition of an asymptotic Killing vector in the electric frame, the conservation law \eqref{WI-e} leads to the conserved charges 
\be\label{conserved-Q-e} 
\wt\cq[\z]= \int_{\partial\cm\cap C} \text{d}^2 \mathbf{x}\,\(2\wt\P^{t}_j+\P^{0t}A^0_j+\wt\p^t\wt A_j^{\rm ren}\)\z^j, 
\ee
which are again manifestly finite. The value of these charges is subject to the same ambiguities as the charges \eqref{conserved-Q}, but as we shall see in the next section, the gauge choice we made in the specification \eqref{subtracted-simple} of the conical backgrounds in the two frames ensures that the charges \eqref{conserved-Q-e} and \eqref{conserved-Q} coincide.

\section{Thermodynamics for asymptotically conical black holes}
\label{thermo}
\setcounter{equation}{0}

In the previous section we derived specific boundary terms that should be added to the STU model action in both the magnetic and electric frames such that the variational problem for asymptotically conical backgrounds of the form \eqref{subtracted-simple} is well posed. Moreover, we showed that the same boundary terms ensure that the on-shell action is free of long-distance divergences and allow us to construct finite conserved charges. In this section we evaluate explicitly these conserved charges and other relevant thermodynamic observables for conical backgrounds and we demonstrate that both the Smarr formula and the first law of thermodynamics hold. Along the way we compare our results with those obtained in \cite{Cvetic:2014nta}, and comment on some differences.  

\subsection{Renormalized thermodynamic observables}

Let us start by evaluating in turn all relevant thermodynamic variables that we will need in order to prove the first law and the Smarr formula. We will use a subscript `4' to denote the variables computed in this section to distinguish them from their counterparts in 5 and 3 dimensions, which we will discuss in section \ref{5&3D}.

\subsubsection*{\small\em Entropy}
The entropy is given by the standard Bekenstein-Hawking area law and its value for the conical black holes \eqref{subtracted-simple} is\footnote{We hope that using the same symbol for the entropy and the action will not cause any confusion, since it should be clear from the context which quantity we refer to.} 
\be\label{entropy}
S_4 =\frac{\pi\ell^2}{G_4}\sqrt{r_+}\;.
\ee

\subsubsection*{\small\em Temperature}
The Hawking temperature can be obtained by requiring that the Euclidean section of the black hole solution is smooth at the horizon, which determines 
\be
\label{temperature}
T_4=\frac{k(r_+ - r_-)}{4\pi\ell\sqrt{r_+}}\;.
\ee

\subsubsection*{\small\em Angular velocity}

We define the physical (diffeomorphism invariant) angular velocity as the difference between the angular velocity at the outer horizon and at infinity, namely  
\be\label{angular-v}
\Omega_4=\Omega_H-\Omega_\infty=\frac{g_{t\f}}{g_{\f\f}}\Big|_{\pa\cm}-\frac{g_{t\f}}{g_{\f\f}}\Big|_{\ch_+}=\o k\sqrt{\frac{r_-}{r_+}}\;.
\ee
In the coordinate system \eqref{subtracted-simple} there is no contribution to the angular velocity from infinity, but there is in the original coordinate system \eqref{subtracted}. The rotation at infinity was not taken into account in \cite{Cvetic:2014nta}, which is why our result does not fully agree with the one obtained there.

\subsubsection*{\small\em Electric charges}

In the magnetic frame there is only one non-zero electric charge given by  
\eqref{e-charges-m}, whose value is 
\be\label{e-charge}
Q_4^{0(e)}=-\int_{\pa\cm\cap C}\text{d}^2 \mathbf{x} \, \P^{0t}=\frac{\ell^4}{4G_4 B^3}\(\sqrt{r_+ r_-}+\o ^2\ell^2\).
\ee
In the electric frame both electric charges defined in \eqref{e-charges-e} are non-zero:
\be
\wt Q_4^{(e)}=-\int_{\pa\cm\cap C}\text{d}^2 \mathbf{x}\;\wt \p^t=\frac{3B}{4G_4}\;,\qquad \wt Q_4^{0(e)}=Q_4^{0(e)}.
\ee

\subsubsection*{\small \em Magnetic charge}

The only non-zero magnetic charge is present in the magnetic frame and it is equal to one of the electric charges in the electric frame:
\bal
\label{m-charges}
Q^{(m)}_4&=-\frac{3}{2\k_4^2}\int_{\pa\cm\cap C} F=\wt Q_4^{(e)}.
\eal

\subsubsection*{\small\em Electric potential}

We define the electric potential as 
\be\label{e-potential}
\F^{0(e)}_4=A^0_i \mathcal{K}^i\Big|_{\ch_+}=k\(\frac B\ell\)^3\sqrt{\frac{r_-}{r_+}}\;, 
\ee
where $\mathcal{K}=\partial_t+\Omega_H\partial_\phi$ is the null generator of the outer horizon. Note that $A^0_i \mathcal{K}^i$ is constant over the horizon  \cite{Papadimitriou:2005ii} and so leads to a well defined electric potential. However, as we remarked in the previous section, the electric potential is not gauge invariant. Under gauge transformations it is shifted by a constant (see \eqref{gauge-parameter}) which compensates the corresponding shift of the charges \eqref{conserved-Q} in the Smarr formula and the first law.

\subsubsection*{\small\em Magnetic potential}

Similarly, the magnetic potential is defined in terms of the gauge field $\wt A_i$ in the electric frame as
\be 
\F^{(m)}_4=\wt A_i \ck^i \Big |_{\ch _+}=\frac{\ell k}{2B}\((r_- -r_+)+2\o ^2\ell^2\sqrt{\frac{r_-}{r_+}} \).
\ee

\subsubsection*{\small\em Mass}

The mass is the conserved charge associated with the Killing vector\footnote{The overall minus sign relative to the Killing vector used in \cite{Papadimitriou:2005ii} can be traced to the fact that the free energy is defined as the Lorentzian on-shell action in section 5 of that paper, while in section 6 it is defined as the Euclidean on-shell action. We adopt the latter definition here.} $\z=-\pa_t-\O_\infty\pa_\f$. Since $\O_\infty=0$ in the coordinate system \eqref{subtracted-simple}, \eqref{conserved-Q} gives\footnote{In \cite{Cvetic:2014nta} the mass for static subtracted geometry black holes was evaluated from the regulated Komar integral and the Hawking-Horowitz prescription and shown to be equivalent. Both the Smarr formula and the first law of thermodynamics were shown to hold in the static case. In the rotating case, the chosen coordinate system of the subtracted metric  in \cite{Cvetic:2014nta} has non-zero angular velocity at spatial infinity which was erroneously  not included in the  thermodynamics analysis of the rotating subtracted geometry. Furthermore, the evaluation of the regulated Komar integral in the rotating  subtracted geometry  would have to be performed; this would lead to an additional contribution to the regulated Komar mass due to rotation, and in turn ensure the validity of the Smarr formula and the first law of thermodynamics. } 
\be\label{mass}
M_4 =-\int_{\partial\cm\cap C} \text{d}^2 \mathbf{x}\, \(2\P^t_{\phantom{t}t}+\P_0^t A^0_t+\P^t A_t\)=\frac{\ell k}{8G_4}(r_+ + r_-)\;.
\ee
The same result is obtained in the electric frame using \eqref{conserved-Q-e}.

\subsubsection*{\small\em Angular momentum}

The angular momentum is defined as the conserved charge corresponding to the Killing vector $\z=\pa_\f$, which gives  
\be\label{ang-momentum}
J_4=\int_{\pa\cm\cap C} \text{d}^2 \mathbf{x}\, (2\P^t_{\phantom{t}\phi}+\P_0^t A^0_\phi + {\P}^t {A}_\phi)=-\frac{\o\ell^3}{2G_4}\;.
\ee
The same result is obtained in the electric frame.

\subsubsection*{\small\em Free energy}

Finally, the full Gibbs free energy, $\wt\cg_4$, is related to the renormalized Euclidean on-shell action in the electric frame, where all charges are electric. Namely, 
\be
\wt I_4=\wt S_{\rm ren}^{\rm E}=-\wt S_{\rm ren}=\b_4 \wt\cg_4,
\ee
with $\b_4=1/T_4$ and $\wt S_{\rm ren}$ defined in \eqref{ren-action-e}. The Euclidean on-shell action in the magnetic frame similarly defines another thermodynamic potential, $\cg_4$, through  
\be
I_4=S_{\rm ren}^{\rm E}=-S_{\rm ren}=\b_4\cg_4,
\ee
where $S_{\rm ren}$ was given in \eqref{ren-action-m}. Evaluating this we obtain (see  appendix \ref{4Daction-evaluation}) 
\be
I_4 =\frac{\b_4\ell k}{8G_4}\((r_- - r_+) +2\o^2\ell^2 \sqrt{\frac{r_-}{r_+}}\).
\ee
Moreover, \eqref{e-frame-action} implies that the on-shell action is given by
\be
\wt I_4= I_4+\frac{3}{2\k_4^2}\int_{\ch_+}\wt A\wedge F=I_4-\b_4\F_4^{(m)}Q_4^{(m)}.
\ee

An interesting observation is that the value of the renormalized action in the magnetic frame, as well as the value of all other thermodynamic variables, is independent of the parameter $\a$ in the boundary counterterms \eqref{counterterms}. This property is necessary in order for the thermodynamic variables in the electric and magnetic frames to agree, and in order to match with those of the 5D uplifted black holes that we will discuss in section \ref{5&3D}. 
Recall that the terms multiplying $\a$ are designed so that their leading asymptotic contribution to the Hamilton-Jacobi solution \eqref{HJ-sol}, as well as to the derivatives \eqref{S-derivatives}, vanishes by means of the asymptotic constraints \eqref{asymptotic-constraints}. This is the reason why any value of $\a$ leads to boundary counterterms that remove the long-distance divergences. However, the parameter $\a$ does appear in the renormalized momenta, as is clear from \eqref{ren-momenta}, and in the unintegrated value of the renormalized action. Nevertheless, $\a$ does not enter in any physical observable. This observation results from the explicit computation of the thermodynamic variables, but we have not been able to find a general argument that ensures this so far.

\subsection{Thermodynamic relations and the first law}

We can now show that the thermodynamic variables we just computed satisfy the expected thermodynamic relations, including the first law of black hole mechanics.   

\subsubsection*{\small\em Quantum statistical relation}

It is straightforward to verify that the total Gibbs free energy $\wt\cg_4$ satisfies the quantum statistical relation \cite{Gibbons:1976ue} 
\be\label{QS1}
\wt\cg_4=M_4 -T_4 S_4 -\O_4 J_4-\F^{0(e)} Q^{0(e)}-\F_4^{(m)}Q_4^{(m)}.
\ee
Similarly, the thermodynamic potential $\cg_4$, which was obtained from the on-shell action in the magnetic frame, satisfies
\be\label{QS2}
\cg_4=M_4 -T_4 S_4 -\O_4 J_4-\F^{0(e)} Q^{0(e)}.
\ee
Note that the shift of the mass and angular momentum under a gauge transformation \eqref{gauge-parameter}  is compensated by that of the electric potentials so that these relations are gauge invariant.

\subsubsection*{\small\em First law}

In order to demonstrate the validity of the first law we must recall the transformations \eqref{var1} and \eqref{var2} of the non-normalizable boundary data that allow for a well posed variational problem. In particular, variations of $B$, $k$ and $\ell$ that are a combination of the two transformations \eqref{var1} and \eqref{var2} are equivalent to generic transformations keeping $kB^3/\ell^3$ fixed. Considering such transformations, as well as arbitrary variations of the normalizable parameters $r_\pm$ and $\o$, we obtain
\be\label{first-law}
\text{d}M_4-T_4 \text{d}S_4 -\O_4 \text{d}J_4-\F^{0(e)}_4 \text{d}Q^{0(e)}_4- \F^{(m)}_4 \text{d}Q^{(m)}_4 =0.
\ee

\subsubsection*{\small\em Smarr formula}

Finally, one can explicitly check that the Smarr formula
\be \label{smarr}
M_4=2S_4 T_4+2\O _4 J_4+Q^{0(e)}_4 \F^{0(e)}_4+Q^{(m)}_4 \F^{(m)}_4,
\ee 
also holds. This identity can be derived by applying the first law to the one-parameter family of transformations  
\be 
\d M_4=\e M_4, \quad \d S_4=2\e S_4,\quad \d J_4=2\e J_4,\quad \d Q^{0(e)}_4=\e Q^{0(e)}_4, \quad \d Q^{(m)}_4=\e Q^{(m)}_4,
\ee 
which corresponds to the parameter variations
\be
\d \ell =\e \ell,\quad \d B=\e B,\quad \d\o=-\e \o,         
\ee
while keeping all other parameters of the solutions fixed. This transformation keeps $kB^3/\ell^3$ fixed and, therefore, it is a special case of the allowed transformations for the variational problem and the first law. The weight of $\o$ under this transformation follows from dimensional analysis.

\section{5D uplift and relation to the BTZ black hole}
\label{5&3D}
\setcounter{equation}{0}

The STU model \eqref{STU} can be obtained by a circle reduction from a five-dimensional theory \cite{Virmani:2012kw}. Kaluza-Klein reducing the resulting theory on an $S^2$ gives rise to Einstein-Hilbert gravity in three dimensions, coupled to several matter fields \cite{Chong:2004na,Virmani:2012kw,Baggio:2012db}. Through this sequence of uplifts and Kaluza-Klein reductions, the conical backgrounds \eqref{subtracted-simple} can be related to the BTZ black hole in three dimensions \cite{Baggio:2012db,Cvetic:2011dn,Cvetic:2014nta}.  

In this section we revisit the uplift of the truncated STU model \eqref{STU-reduced} to five dimensions, as well as the reduction of the resulting 5D theory to three dimensions, keeping track of all surface terms on the boundary and on the horizon. As we will demonstrate, these terms are essential in order to connect the thermodynamics of the 4D black holes with that of the BTZ black hole. Moreover, we find that some continuous parameters of the 4D solutions must be quantized in order for the uplift to 5D to be possible, which explains the mismatch between the number of thermodynamic variables in four and three dimensions.

\subsection{4D action from circle reduction}

A consistent truncation of the 5D uplift of the STU model is given by the action
\cite{Virmani:2012kw,Baggio:2012db}
\be\label{5D-action}
S_5=\frac{1}{2\k_5^2}\int_{\Hat\cm}\text{d}^5 \mathbf{x} \(R[\Hat g]\star1-\frac{3}{2}\star\Hat{F}\wedge\Hat{F}+\Hat{F}\wedge\Hat{F}\wedge\Hat{A}\)+\frac{1}{2\k_5^2}\int_{\pa\Hat\cm}\text{d}^4\mathbf{x}\;\sqrt{-\Hat \g}\;2K[\Hat\g],
\ee
where hats signify 5D quantities. If $z$ is a compact dimension of length $R_z$, then the Kaluza-Klein ansatz
\be\label{reduction-ansatz}
d\Hat s^2=e^\h ds^2+e^{-2\h} (dz+A^0)^2,\qquad \Hat A=\c(dz+A^0)+A, 
\ee
gives \cite{Pope:lectures}
\begin{subequations}
\bal
\label{5DRicci}
\sqrt{-\Hat g}\;R[\Hat g] & = \sqrt{-g}\left(R[g]-\frac{3}{2}\pa_{\mu}\h\pa^{\mu}\h-\frac{1}{4}e^{-3\h}F_{\mu\nu}^{0}F^{0\mu\nu}-\square_g\h\right),\\
\label{5DGaugeKinetic}
\sqrt{-\Hat g}\;\frac{1}{4}\Hat{F}^{2}&=\sqrt{-g}\( \frac{1}{4}e^{-\h}(F+\c F^{0})_{\mu\nu}(F+\c F^{0})^{\mu\nu}+\frac{1}{2}e^{2\h}\pa_{\mu}\c\pa^{\mu}\c\),\\
\Hat F\wedge \Hat F\wedge \Hat A   &=  dz\wedge\(3\c F\wedge  F+3\c^2 F\wedge F^0 +\c^3F^0\wedge F^0 -d\(\c^2  A\wedge F^0+2\c  A\wedge F\)\).
\eal
\end{subequations}

In order to reduce the Gibbons-Hawking term we need the canonical decomposition \eqref{ADM} of the 5D metric, which takes the form
\be
d\Hat s^2=\Hat N^2 du^2+\Hat\g_{\hat i\hat j}dx^{\hat i}dx^{\hat j}=e^\h N^2du^2+e^\h\g_{ij}dx^idx^j+e^{-2\h}(dz+A^0_idx^i)^2,
\ee
where $\hat i=(z,i)$. In matrix form, therefore, the induced metric, $\Hat\g_{\hat i\hat j}$, on the four-dimensional radial slices $\Hat\S_u$ is related to the induced fields on the three-dimensional radial slices $\S_u$ via
\be
\Hat\g_{\hat i\hat j}=\(\begin{matrix}
e^{-2\h} & e^{-2\h} A_i^0 \\
e^{-2\h} A_i^0 & e^\h \g_{ij}+e^{-2\h} A_i^0A_j^0\\
\end{matrix}\),\qquad 
\Hat\g^{\Hat i\Hat j}=\(\begin{matrix}
e^{2\h}+e^{-\h}A^0_kA^{0k} & -e^{-\h} A^{0i} \\
-e^{-\h} A^{0i} & e^{-\h} \g^{ij}\\
\end{matrix}\).
\ee
From these expressions it is straightforward to compute $\det\Hat\g=e^\h\det\g$.  Moreover, the extrinsic curvature of $\Hat\g_{\hat i\hat j}$ is given by 
\be
K[\Hat\g]_{\hat i\hat j}=\frac{1}{2\Hat N}\dot{\Hat \g}_{\hat i\hat j},
\ee
and can be expressed in terms of four-dimensional variables as 
\bsub
\bal
K[\Hat\g]_{zz}&=-\frac1N e^{-5\h/2}\dot \h,\\
K[\Hat\g]_{zi}&=-\frac{1}{N} e^{-5\h/2}\(\dot \h A^0_i-\frac12\dot A_i^0\),\\
K[\Hat\g]_{ij}&=\frac{1}{N}e^{-\h/2}\(\frac12e^\h\dot\h\g_{ij}+e^{-2\h}A_{(i}^0\dot A_{j)}^0-e^{-2\h}\dot\h A_i^0A_j^0+\frac12 e^\h\dot\g_{ij}\).
\eal
\esub
In particular, the trace of the extrinsic curvature is given by
\be
K[\Hat\g]=\Hat\g^{\hat i\hat j}K[\Hat\g]_{\hat i\hat j}=\frac{1}{2N}e^{-\h/2}\dot\h+e^{-\h/2}K[\g],
\ee
which allows us to reduce the 5D Gibbons-Hawking term to 4D. 

Combining the reduction formulae for the bulk and Gibbons-Hawking terms leads to the four-dimensional action
\be\label{5Dto4Dreduction}
S_5=S_4-\frac{1}{2\k_4^2}\int_{\pa\cm}\(\c^2  A\wedge F^0+2\c A\wedge  F\)+\frac{1}{2\k_4^2}\int_{\ch_+}\(\c^2  A\wedge F^0+2\c A\wedge  F\),
\ee
where $S_4$ is the magnetic frame action \eqref{STU-reduced}, and the 5D and 4D gravitational constants are related via $\k_5^2=R_z\k_4^2$. Hence, even though the 4D magnetic frame action can be obtained by a circle reduction from the 5D action \eqref{5D-action}, there are additional surface terms that are necessary for connecting the physics in 4 and 5 dimensions. In particular, the surface term on the boundary vanishes on-shell when evaluated on the conical backgrounds \eqref{subtracted-simple}, but it is required in order to properly relate the 5D and 4D variational problems. Moreover, the surface term on the horizon is necessary to relate the free energies. 

However, we also need to uplift the boundary counterterms \eqref{counterterms} so that the five-dimensional on-shell action is free of long-distance divergences and the variational problem is well posed. Since   
\bsub
\bal
\sqrt{-\Hat\g} &=\sqrt{-\g} \;e^{\h/2},\\
\sqrt{-\Hat\g}\;R[\Hat\g]&=\sqrt{-\g}\;e^{-\h/2}\(R[\g]-\frac18\pa_i\h\pa^i\h-\frac14e^{-3\h}F_{ij}^0F^{0ij}\)+\text{total derivative},\\
\sqrt{-\Hat\g}\;\Hat F_{ij}\Hat F^{ij}&=\sqrt{-\g}\;e^{-\h/2}\(e^{-\h} F_{ij}F^{ij}-2e^{2\h}\pa_i\c\pa^i\c\),
\eal
\esub
it follows that the boundary counterterms for the 4D action can be uplifted to five dimensions provided they are a linear combination of the expressions on the RHS of these identities. Moreover, the same counterterms must coincide with \eqref{counterterms} up to finite local counterterms and at least for some specific value of the parameter $\a$, or else the variational problem in four dimensions would not be well defined. The only way to reconcile these conditions is by setting $\a=0$ in \eqref{counterterms} and adding the finite local counterterm $\sqrt{-\g}e^{-\h/2}(\pa\h)^2$ with the appropriate coefficient.\footnote{Note that the value of the parameter $\a$ required for the uplift to 5D ($\a=0$) is different from that required in the electric frame ($\a=3$). This reflects the fact that the variational problems in the two cases are somewhat different, with the uplift to 5D only being possible provided $B$ is kept fixed and $\o$ is quantized in units of $1/2Bk$, as we will see below.} The resulting boundary counterterms are  
\bal\label{counterterms-5D} 
S_{\rm ct}'&=-\frac{1}{\k_4^2}\int\text{d}^3 \mathbf{x}\sqrt{-\g}\;\frac{1}{B}e^{\h/2}\(1-\frac14 B^2 e^{-\h}R[\g]+\frac{1}{16}B^2 e^{-4\h}F^0_{ij}F^{0ij}+\frac{1}{32}B^2e^{-\h}\pa_i\h\pa^i\h\), 
\eal
whose uplift is
\be\label{5DCounterterms} 
S_{\rm ct}'=-\frac{1}{\k_5^2}\int\text{d}^4 \mathbf{x}\sqrt{-\Hat\g}\;\frac{1}{B}\(1-\frac14 B^2R[\Hat\g]\). 
\ee

\subsection{Uplifting conical backgrounds to 5D}

Uplifting the conical black hole solutions \eqref{subtracted-simple} using the Kaluza-Klein ansatz \eqref{reduction-ansatz} results in the 5D background \cite{Cvetic:2014ina}
\bsub
\bal
d\Hat s^2&=\frac{4B^2 \r^2 d\r^2}{(\r^2-\r^2_+)(\r^2-\r^2_-)}-\frac{(\r^2-\r^2_+)(\r^2-\r^2_-)}{4B^2\r^2}d t^2+\r^2\(d\f_3-\frac{\r_+ \r_-}{2B\r^2}d t\)^2\NO\\
&\hskip0.2in+B^2\(d\th^2+\sin^2\th\(d\f+2Bk\o d\f_3\)^2\),\label{5Dmetric}\\
\Hat A&=B\cos\th\(d\f+2Bk\o d\f_3\),\label{5DGauge}
\eal
where the new coordinates $\r$ and $\f_3$ are defined through the relations
\esub
\be\label{BTZ-coords}
z=2B k\(\frac{B}{\ell}\)^3 \f_3,\qquad r =\frac{1}{(2Bk)^2} \(\frac{B}{\ell}\)^{-2}\r^2.
\ee
In this coordinate system the 5D metric \eqref{5Dmetric} is immediately recognizable as a 2-sphere of radius $B$, fibered over a three-dimensional BTZ black hole \cite{Banados:1992wn} with AdS$_3$ radius $L=2B$. Since the BTZ angular coordinate $\f_3$ must have periodicity $2\p$, the length $R_z$ of the 5D circle is determined through \eqref{BTZ-coords} to be 
\be 
R_z=4\p B k\(\frac{B}{\ell}\)^3. 
\ee
Given that the gravitational constants in four and five dimensions are related by $\k_5^2=R_z\k_4^2$, this implies that the variational problem in five dimensions must be formulated keeping $B$ fixed, in addition to $kB^3/\ell^3$, which must be kept fixed even in four dimensions. Moreover, the internal $S^2$ has a conical singularity at the north and south poles unless $2Bk\o$ is an integer. This implies that the conical backgrounds \eqref{subtracted-simple} can be uplifted to five dimensions if and only if $\o$ is quantized in units of $1/(2Bk)$. With this condition, the internal part of the metric \eqref{5Dmetric} becomes the standard metric on $S^2$ with azimuthal coordinate $\f'=\f+n\f_3$, where $n\in \bb Z$.

\subsection{$S^2$ reduction and BTZ thermodynamics}

The 5D action \eqref{5D-action} can be Kaluza-Klein reduced on the internal $S^2$ using the reduction ansatz \cite{Baggio:2012db}
\be
\label{3Dansatz}
d\Hat s^2=ds_3^2+B^2d\O^2_2,\qquad \Hat A=B\cos\th\(d\f+2Bk\o d\f_3\).
\ee
The resulting theory in three dimensions is Einstein-Hilbert gravity
\be
S_5=S_3=\frac{1}{2\k_{3}^{2}}\left(\int_{\cm_3}\text{d}^3\mathbf{x}\sqrt{-g_{3}}(R_{3}-2\L_3)+\int_{\pa\cm_3}\text{d}^2\mathbf{x}\sqrt{-\g_{2}}\; 2K_{2}\right),\label{3Daction}
\ee
with cosmological constant $\L_3=-1/(2B)^2$ and gravitational constant given by
\be\label{Newtons}
\k_3^2=\frac{\k_5^2}{(2B)^2\p}=\frac{\k_4^2}{B}k\(\frac{B}{\ell}\)^3.
\ee 

Moreover, from \eqref{3Dansatz} follows that 
\be
R[\Hat\g]=\frac{2}{B^2}+R[\g_2],
\ee
and so the boundary counterterms \eqref{5DCounterterms} for the five-dimensional theory reduce in three dimensions to the boundary terms 
\be\label{3D-counterterms}
S_{\rm ct}'=-\frac{1}{\k_3^2}\int\text{d}^2 \mathbf{x}\sqrt{-\g_2}\;\(\frac{1}{2B}-\frac B4R[\g_2]\).
\ee
The first term is the standard volume divergence of an AdS$_3$ space with radius $L=2B$. The second term is proportional to the Euler density of the induced metric $\g_2$ and corresponds to a particular renormalization scheme. It shifts the on-shell action by a finite multiple of the Euler characteristic of the AdS$_3$ boundary. However, three-dimensional solutions with non-trivial $z$ dependence, such as those obtained by turning on a generic metric source $\g_2$ on the AdS$_3$ boundary, excite Kaluza-Klein fields in the circle reduction to 4D and, therefore, are not captured by the STU model. 4D solutions of the STU model uplift to 5D solutions that are oxidized along the $z$ coordinate, and consequently reduce to 3D solutions that can only have a non-trivial profile along an AdS$_2$ inside the AdS$_3$. For such solutions $R[\g_2]$ vanishes identically, which explains why the boundary counterterms \eqref{3D-counterterms} we obtained from the STU model do not include the logarithmic counterterm $-\frac B2 R[\g_2]\log\e^2$ corresponding to the conformal anomaly of the dual CFT$_2$ \cite{Henningson:1998gx}. 

Combining \eqref{5Dto4Dreduction} and \eqref{3Daction}, the renormalized action in three dimensions can be related to that of the STU model in the magnetic frame, namely
\be\label{3Dto4D}
S_3+S_{\rm ct}'=S_4+S_{\rm ct}'+\frac{1}{2\k_4^2}\int_{\ch_+}\(\c^2  A\wedge F^0+2\c A\wedge  F\),
\ee
where we have used the fact that the surface term on the boundary in \eqref{5Dto4Dreduction} vanishes identically for the conical solutions \eqref{subtracted-simple}. However, the contribution on the horizon is non-zero, which implies that the value of the Gibbs free energies in three and four dimensions do not coincide. More specifically, the complete set of relations between the BTZ thermodynamic variables \cite{Banados:1992wn}
\bal
\label{3Dvariables}
\begin{aligned}
& T_{3}  =  \frac{\r_{+}^{2}-\r_{-}^{2}}{2\pi L^{2}\r_{+}},\quad S_{3} = \frac{4\pi^{2}\r_{+}}{\kappa_{3}^{2}},\quad M_3=\frac{\p}{\k_3^2L^2}\(\r_+^2 +\r_-^2\),  \quad \\
&  \Omega_{3} = \frac{\r_{-}}{L\r_{+}},\quad J_3=\frac{2\pi \r_{+}\r_{-}}{\k_{3}^{2}L},\quad I_3=\frac{\pi\beta_{3}}{\kappa_{3}^{2}L^{2}}(\r_{-}^{2}-\r_{+}^{2}),
\end{aligned}
\eal 
and the 4D ones computed in section \ref{thermo} is
\bal
\label{4Dvs3D}
\begin{aligned}
&T_4=T_3,\qquad S_4=S_3,\qquad M_4=M_3,\qquad I_4+\frac12\b_4\O_4J_4=I_3,\\
&\O_4=(2Bk\o)\O_3=n\O_3,\quad n\in \bb Z,\qquad J_4=-(2Bk\o)\frac{\p L}{\k_3^2}=-\frac{n\p L}{\k_3^2},\\
&\F_4^{0(e)}=Lk\(\frac{B}{\ell}\)^3\O_3,\qquad \F_4^{0(e)} Q_4^{0(e)}+\frac12\O_4J_4=\O_3J_3,  \\
&\F_4^{(m)}Q_4^{(m)}+\frac32\O_4J_4=-\frac32T_3S_3,\qquad Q_4^{(m)}=\frac{6\p}{\k_3^2}k\(\frac B\ell\)^3.
\end{aligned} 
\eal
Clearly, besides the mass, entropy and temperature, the relation between the 3D and 4D variables is non-trivial. In particular, the 3D thermodynamics ensemble corresponds to a subspace of the 4D ensemble, since the 4D angular momentum and magnetic charge are fixed constants in the 5D and 3D thermodynamics, 
which also renders the corresponding potentials $\O_4$ and $\F^{(m)}_4$ redundant. This is a direct consequence of the fact that the magnetic field $B$ must be kept fixed in the 5D and 3D variational problems, while the rotation parameter $\o$ must be quantized in units of $1/(2Bk)$.  

We end this section with the observation that inserting the relations \eqref{4Dvs3D} into the 4D quantum statistical relation \eqref{QS2} and the first law \eqref{first-law} we obtain the corresponding 3D thermodynamic identities, namely   
\be
I_{3}=\beta_{3}(M_{3}-T_{3}S_{3}-\Omega_{3}J_{3}),
\ee
and 
\be\label{3D-first-law}
\text{d}M_{3}=T_{3}\text{d}S_{3}+\O_{3}\text{d}J_{3}.
\ee
The fact that $J_4$ and $Q_4^{(m)}$ must be kept fixed in the 3D variational problem is crucial for deriving the first law in three dimensions from its 4D counterpart. Moreover, the Smarr formula \eqref{smarr} gives
\be
M_3=\frac12T_3S_3+\O_3J_3,
\ee
which can be verified explicitly from the expressions \eqref{3Dvariables}. 
This identity follows from the scaling transformation $\d M_3=2\e M_3$, $\d J_3=2\e J_3$, $\d S_3 =\e S_3$, corresponding to rescaling the BTZ parameters according to $\r_\pm\to (1+\e)\r_\pm$.

\section{Concluding remarks}
\label{conclusion}
\setcounter{equation}{0}

The main message we would like to get across in this paper is that a well defined thermodynamics, including finite conserved charges and thermodynamic identities, is an immediate consequence of a well posed variational problem, formulated in terms of equivalence classes of boundary data under the asymptotic local symmetries of the theory. This has been known for some time in the case of asymptotically AdS black holes, but we argue that it applies to more general asymptotics, including cases where matter fields are required to support the background. 

We demonstrated this claim by carefully analyzing the variational problem for asymptotically conical backgrounds of the STU model in four dimensions and deriving the thermodynamics of subtracted geometry black holes. Moreover, by uplifting these solutions to five dimensions, we provided a precise map between all thermodynamic variables of subtracted geometries and those of the BTZ black hole. Crucial to this matching was the fact that some free parameters of the four-dimensional black holes must be fixed or quantized in order for the solutions to be uplifted to five dimensions.  

Although our analysis here does not assume or imply a holographic duality for asymptotically conical backgrounds, we would like to view it as a first step in this direction. Our comparison of the variational problems in four and five dimensions indicates that not all asymptotically conical solutions of the STU model in four dimensions correspond to asymptotically AdS$_3\times S^2$ solutions in five dimensions and vice versa. This suggests that the Hilbert space of a putative holographic dual to subtracted geometries can at most have a partial overlap with that of the two-dimensional CFT at the boundary of AdS$_3$. The next steps in order to construct a genuine dual to asymptotically conical backgrounds, as well as to understand the connection with the two-dimensional CFT, would be a systematic analysis of the most general asymptotically conical solutions of the STU model (i.e. not merely stationary), and the identification of the symmetry algebra acting on the modes as a result of the asymptotic local symmetries. We plan to address both these problems in future work.

\section*{Acknowledgments}

O.A. would like to thank Jin U Kang for interesting discussions. M.C. would like to thank Gary Gibbons and Zain Saleem for discussions on related topics. M.C. and I.P. would like to thank the International Institute of Physics, Natal, for the hospitality and financial support during the School on Theoretical Frontiers in Black Holes and Cosmology, June 8--19, 2015, where part of this work was carried out. This research is supported in part by the DOE Grant Award DE-SC0013528, (M.C.), the Fay R. and Eugene L. Langberg Endowed Chair (M.C.) and the Slovenian Research Agency (ARRS) (M.C.).

\appendix

\renewcommand{\thesection}{\Alph{section}}
\renewcommand{\theequation}{\Alph{section}.\arabic{equation}}

\section*{Appendices}
\setcounter{section}{0}

\section{Radial Hamiltonian formalism}
\label{ham}
\setcounter{equation}{0}

In this appendix we present in some detail the radial Hamiltonian formulation of the reduced STU $\s$-model \eqref{STU-reduced}. This analysis can be done abstractly, without reference to the explicit form of the $\s$-model functions $\cg^{IJ}$, $\cz_{\L\S}$ and $\car_{\L\S}$, and it therefore applies to the electric Lagrangian \eqref{STU-reduced-e} as well, provided $A_L$, $\cz_{\L\S}$ and $\car_{\L\S}$ are replaced with their electric frame analogues in \eqref{e-frame}.  

The first step towards a Hamiltonian formalism is picking a suitable radial coordinate $u$ such that constant-$u$ slices, which we will denote by $\S_u$, are diffeomorphic to the boundary $\pa\cm$ of $\cm$. Moreover, it is convenient to choose $u$ to be proportional to the geodesic distance between any fixed point in $\cm$ and a point in $\S_u$, such that\footnote{We assume $\cm$ to be a non-compact space with infinite volume such that the geodesic distance between any point in the interior of $\cm$ and a point in $\pa\cm$ is infinite.} $\S_u\to \pa\cm$ as $u\to\infty$. Given the radial coordinate $u$, we then proceed with an ADM-like decomposition of the metric and gauge fields \cite{Arnowitt:1960es} 
\bal\label{ADM}
ds^2&=(N^2+N_iN^i)du^2+2N_idudx^i+\g_{ij}dx^i dx^j,\NO\\
A^L&=a^\L du+A^\L_idx^i,
\eal
where $\{x^i\}=\{t,\th,\f\}$. This is merely a field redefinition, trading the fully covariant fields $g_{\m\n}$ and $A^L_\m$ for the induced fields $N$, $N_i$, $\g_{ij}$, $a^\L$ and $A^\L_i$ on $\S_u$. Inserting this decomposition in the $\s$-model action \eqref{STU-reduced} and adding the Gibbons-Hawking term \eqref{GH} leads to the radial Lagrangian
\begin{align}\label{r-lagrangian}
L=\frac{1}{2\k^2_4}\int \text{d}^3 \mathbf{x}N\sqrt{-\g}&\left\{R[\g]+K^2-K_{ij}K^{ij}-\frac{1}{2N^2}\cg_{IJ}(\vf)\left(\dot\vf^I-N^i\pa_i\vf^I\right)\left(\dot\vf^J-N^j\pa_j\vf^J\right)\right. \NO\\
&\left.-\frac{2}{N^2}\cz_{\L\S}(\vf)\g^{ij}\left(\dot A_i^\L-\pa_ia^\L-N^kF^\L_{ki}\right)\left(\dot A_j^\S-\pa_ja^\S-N^lF^\S_{lj}\right)\right. \\
&\left.-4\car_{\L\S}(\vf)\e^{ijk}\left(\dot A^\L_i-\pa_ia^\L\right)F^\S_{jk}-\frac12\cg_{IJ}(\vf)\pa_i\vf^I\pa^i\vf^J-\cz_{\L\S}(\vf)F^\L_{ij}F^{\S ij}\right\},\NO
\end{align}
where
\be
K_{ij}=\frac{1}{2N}\left(\dot{\g}_{ij}-D_iN_j-D_jN_i\right),
\ee
is the extrinsic curvature of the radial slices $\S_u$, $D_i$ denotes a covariant derivative with respect to the induced metric $\g_{ij}$ on $\S_u$, while a dot $\dot{}$ stands for a derivative with respect to the Hamiltonian `time' $u$. 

The canonical momenta conjugate to the induced fields on $\S_u$ following from the Lagrangian (\ref{r-lagrangian}) are 
\begin{subequations}\label{momenta}
\begin{align}
\pi^{ij}=&\frac{\d L}{\d\dot{\g}_{ij}}=\frac{1}{2\k_4^2}\sqrt{-\g}\left(K\g^{ij}-K^{ij}\right), \\
\pi_I=&\frac{\d L}{\d\dot\vf^I}=-\frac{1}{2\k_4^2}N^{-1}\sqrt{-\g}\;\cg_{IJ}\left(\dot\vf^J-N^i\pa_i\vf^J\right), \\
\pi^i_\L=&\frac{\d L}{\d\dot A^\L_i}=-\frac{2}{\k_4^2}N^{-1}\sqrt{-\g}\cz_{\L\S}\left(\g^{ij}\left(\dot A^\S_j-\pa_ja^\S\right)-N_jF^{\S ji}\right)-\frac{2}{\k_4^2}\sqrt{-\g}\;\car_{\L\S}\e^{ijk}F^\S_{jk}.
\end{align}
\end{subequations}
Notice that the momenta conjugate to $N$, $N_i$, and $a^\L$ vanish identically, since the Lagrangian (\ref{r-lagrangian}) does not contain any radial derivatives of these fields. It follows that the fields $N$, $N_i$, and $a^\L$ are Lagrange multipliers, implementing three first class constraints, which we will derive momentarily. The canonical momenta (\ref{momenta}) allow us to perform the Legendre transform of the Lagrangian (\ref{r-lagrangian}) to obtain the radial Hamiltonian
\be\label{H}
H=\int \text{d}^3 \mathbf{x}\left(\pi^{ij}\dot{\g}_{ij}+\pi_I\dot\vf^I+\pi_\L^i\dot A^\L_i\right)-L=\int \text{d}^3 \mathbf{x}\left(N\mathcal{H}+N_i\ch^i+a^\L\cf_\L\right),
\ee
where
\begin{subequations}\label{constraints}
\begin{align}
\ch=&-\frac{\k^2_4}{\sqrt{-\g}}\left(2\left(\g_{ik}\g_{jl}-\frac{1}{2}\g_{ij}\g_{kl}\right)\pi^{ij}\pi^{kl}+\cg^{IJ}(\vf)\p_I\p_J\right.\NO\\
&\left.+\frac{1}{4}\cz^{\L\S}(\vf)\left(\pi_{\L i}+\frac{2}{\k^2_4}\sqrt{-\g}\car_{\L M}(\vf)\e_i{}^{kl}F^M_{kl}\right)\left(\pi^i_\S+\frac{2}{\k^2_4}\sqrt{-\g}\car_{\S N}(\vf)\e^{ipq}F^N_{pq}\right)\right) \NO \\
&+\frac{\sqrt{-\g}}{2\k^2_4}\left(-R[\g]+\frac12\cg_{IJ}(\vf)\pa_i\vf^I\pa^i\vf^J+\cz_{\L\S}(\vf)F^\L_{ij}F^{\S ij}\right), \\
\ch^i=&-2D_j\pi^{ij}+\pi_I\pa^i\vf^I+F^{\L ij}\left(\pi_{\L j}+\frac{2}{\k^2_4}\sqrt{-\g}\car_{\L\S}(\vf)\e_j{}^{kl}F^\S_{kl}\right), \\
\cf_\L=&-D_i\pi^i_\L.
\end{align}
\end{subequations}
Since the canonical momenta conjugate to the fields $N$, $N_i$, and $a^\L$ vanish identically, the corresponding Hamilton equations lead to the first class constraints
\be\label{1st-class-constraints}
\ch=\ch^i=\cf_\L=0,
\ee
which reflect respectively diffeomorphism invariance under radial reparameterizations, diffeomorphisms along the radial slices $\S_u$ and a $U(1)$ gauge invariance for every gauge field $A_i^\L$.

\subsection*{Hamilton-Jacobi formalism}

The first class constraints \eqref{1st-class-constraints} are particularly useful in the Hamilton-Jacobi formulation of the dynamics, where the canonical momenta are expressed as gradients of Hamilton's principal function $\cs[\g,A^\L,\vf^I]$ as 
\be\label{HJ-momenta-explicit}
\p^{ij}=\frac{\d\cs}{\d\g_{ij}},\quad \p_\L^i=\frac{\d\cs}{\d A^\L_i},\quad \p_I=\frac{\d\cs}{\d\vf^I}.
\ee
Since the momenta conjugate to $N$, $N_i$, and $a^\L$ vanish identically, the functional $\cs[\g,A^\L,\vf^I]$ does not depend on these Lagrange multipliers. Inserting the expressions \eqref{HJ-momenta-explicit} for the canonical momenta in the first class constraints \eqref{1st-class-constraints} leads to a set of functional partial differential equations for $\cs[\g,A^\L,\vf^I]$. These are the Hamilton-Jacobi equations for the Lagrangian \eqref{r-lagrangian}. 

Given a solution $\cs[\g,A^\L,\vf^I]$ of the Hamilton-Jacobi equations, the radial evolution of the induced fields $\g_{ij}$, $a^\L$ and $A^\L_i$ is determined through the first order equations obtained by identifying the expressions \eqref{momenta} and \eqref{HJ-momenta-explicit} for the canonical momenta. Namely, gauge-fixing the Lagrange multipliers $N_i=a^\L=0$, but keeping $N$ arbitrary, the resulting first order equations are 
\begin{subequations}\label{flow}
\begin{align}
\frac1N\dot{\g}_{ij}=&-\frac{4\k^2_4}{\sqrt{-\g}}\left(\g_{ik}\g_{jl}-\frac{1}{2}\g_{ij}\g_{kl}\right)\frac{\d \cs}{\d\g_{kl}}, \\
	\frac1N\dot{\vf}^I=&-\frac{2\k^2_4}{\sqrt{-\g}}\cg^{IJ}(\vf)\frac{\d \cs}{\d\vf^J} ,\\
\frac1N	\dot{A}_i^\L=&-\frac{\k^2_4}{2\sqrt{-\g}}\cz^{\L\S}(\vf)\g_{ij}\frac{\d \cs}{\d A_j^\S}-\cz^{\L\S}(\vf)\car_{\S P}(\vf)\e_i\,^{jk}F^P_{jk}.
\end{align}
\end{subequations}
The complete solution of the equations of motion can be obtained by solving the Hamilton-Jacobi equations, together with the first order equations \eqref{flow}, without actually solving the second order equations of motion. Even though this may not seem an easier avenue to solve the system, it is a very efficient approach for obtaining asymptotic solutions of the equations of motion, which is all that is required in order to determine the boundary terms that render the variational problem well posed \cite{Papadimitriou:2010as}.  

These boundary terms, commonly referred to as `boundary counterterms', can in fact be read off a suitable asymptotic solution $\cs[\g,A^\L,\vf^I]$ of the Hamilton-Jacobi equations \cite{Papadimitriou:2010as}. This is related to the fact that Hamilton's principal function generically coincides with the on-shell action,\footnote{This holds provided the Hamilton's principal function in question corresponds, through the first order equations \eqref{flow}, to asymptotic solutions satisfying the same boundary conditions as the solutions on which the action is evaluated.} up to terms that remain finite as $\S_u\to\pa\cm$. In particular, the divergent part of $\cs[\g,A^\L,\vf^I]$ coincides with that of the on-shell action. Adding, therefore, the boundary counterterms $S_{\rm ct}=-\cs$ to the action, where $\cs[\g,A^\L,\vf^I]$ is a suitable asymptotic solution of the Hamilton-Jacobi equations, not only renders the variational problem well posed, but also automatically ensures that the on-shell action remains finite as $\S_u\to\pa\cm$ \cite{Papadimitriou:2005ii,Papadimitriou:2010as}. For asymptotically AdS backgrounds, the fact that the divergences of the on-shell action can be canceled by a solution of the radial Hamilton-Jacobi equation was first observed in \cite{deBoer:1999tgo}.

\section{Evaluation of the 4D renormalized on-shell action}
\label{4Daction-evaluation}
\setcounter{equation}{0}

The easiest way to evaluate the renormalized on-shell action of the reduced STU model in the magnetic frame is to utilize the relation \eqref{3Dto4D}, namely
\be 
S_{\rm ren}=\lim_{r\to\infty}\(S_4+S_{\rm ct}^\prime\) =\lim_{r\to\infty}\(S_3+S_{\rm ct}^\prime\)-\frac{1}{2\k _4^2}\int_{\ch _+} \(\c ^2 A\wedge F^0+2\c A\wedge F \),
\ee 
which relates $S_{\rm ren}$ to the renormalized on-shell action in three dimensions, plus a surface contribution from the outer horizon. The renormalized on-shell action in three dimensions is (see \eqref{3Dvariables}) 
\be 
\lim_{r\to\infty}\(S_3+S_{\rm ct}^\prime\) =\frac{\p \b _3}{\k _3^2 L^2}\(\r _+^2-\r _-^2 \)=\frac{\b_4k\ell}{8G_4}(r_+-r_-),
\ee 
where \eqref{BTZ-coords} and \eqref{Newtons} have been used in the second step. Moreover, the parity-odd term on the horizon gives
\be
\frac{1}{2\k _4^2}\int_{\ch _+} \(\c ^2 A\wedge F^0+2\c A\wedge F \)=\frac{k\ell^3}{2\k_4^2}\int \text{d}^3\mathbf{x}\ \pa_\th \(\frac{\o^2\sqrt{r_+ r_-}\cos^3\th}{r_+ +\o^2\ell^2\sin^2\th}\)=-\frac{\b _4}{4G_4}k\ell^3\o ^2\sqrt{\frac{r_-}{r_+}}.
\ee 
Combining these two results we obtain 
\be \label{ren-onshell-action}
S_{\rm ren}=\frac{k\ell }{8G_4}\(r_+-r_--2\ell^2\o^2\sqrt{\frac{r_-}{r_+}} \).
\ee 

A few comments are in order here. Firstly, although in the gauge in which the backgrounds \eqref{subtracted-simple} are given the parity-odd terms on the boundary in \eqref{5Dto4Dreduction} give a zero contribution, this is not the case for a generic choice of gauge for the potential $A$. In general both contributions from the boundary and the horizon must be considered, and their difference is clearly gauge invariant.   

A second comment concerns the potential dependence of the renormalized on-shell action on the parameter $\a$. Here we have evaluated the renormalized on-shell action through the relation \eqref{3Dto4D}, which holds only for $\a =0$. However, evaluating the counterterms \eqref{counterterms} for generic $\a $ we obtain
\be 
S_{\rm ct}=-\frac{\ell}{\k_4^2}\int_{\cm _{r_0}}\text{d}^3\mathbf{x}\;\sin\th\(\half r_0-\frac{1}{4}(r_+ + r_-)-\frac\a8\o^2\ell^2(1+3\cos 2\th)+\co\(r_0^{-1}\) \),
\ee 
where $r_0$ is the radial cut-off. It is obvious that the $\a $-dependent term drops out after integration over $\th$, which implies that for all values of $\a $ we get the same result \eqref{ren-onshell-action}. Therefore, the renormalized on-shell action is independent of the choice of $\a$. 

The same conclusion holds for the finite counterterm $\sqrt{-\g}e^{-\h/2}(\pa\h)^2$ that was added in \eqref{counterterms-5D} in order to uplift the counterterms to five dimensions. Namely, this term does not contribute to the on-shell action since 
\be 
\int \text{d}^3\mathbf{x}\sqrt{-\g}e^{\h/2}\(e^{-\h}\pa _i \h \pa ^i\h \)=\int \text{d}^3\mathbf{x} \;\frac{k\o ^2\ell^3}{B}\(\sin^3\th -2\cos^2\th \sin\th  \)=0.
\ee 
Hence, evaluating $S_{\rm ren}$ with $S_{\rm ct}'$ in \eqref{counterterms-5D} or with $S_{\rm ct}$ in \eqref{counterterms} gives the same result \eqref{ren-onshell-action}.

\addcontentsline{toc}{section}{References}


\bibliographystyle{jhepcap}
\bibliography{subtracted}

\end{document}